\documentclass[a4paper,fleqn,usenatbib]{mnras}
\usepackage{newtxtext,newtxmath}
\usepackage[T1]{fontenc}
\usepackage{ae,aecompl}
\usepackage{graphicx}	% Including figure files
\usepackage{ulem}
\usepackage{amsmath}	% Advanced maths commands
\usepackage{amssymb}	% Extra maths symbols
\usepackage{color}      
\definecolor{darkgreen}{rgb}{0,0.7,0}
\definecolor{magen}{rgb}{0.79,0.08,0.48}

\newcommand{\grale}{\textsc{Grale}}
\newcommand{\lenstool}{\textsc{Lenstool}}

\newcommand{\as}{$^{\prime\!\prime}$}

\newcommand{\bfit}{BestFitness}
\newcommand{\hres}{HighResolution}

\newcommand{\simgt}{\hbox{\,\rlap{\raise 0.425ex\hbox{$>$}}\lower 0.65ex\hbox{$\sim$}\,}}
\newcommand{\simlt}{\hbox{\,\rlap{\raise 0.425ex\hbox{$<$}}\lower 0.65ex\hbox{$\sim$}\,}}

\renewcommand{\vec}[1]{\boldsymbol{#1}} % This should replace the arrows in vector notation with a bold symbol

\title[Measuring $H_0$ from SN Refsdal in MACS J1149]
{The role of multiple images and model priors in measuring $H_0$ from supernova Refsdal in galaxy cluster MACS J1149.5+2223}

\author[L.L.R. Williams et al.]{
Liliya L.R. Williams,$^{1}$\thanks{E-mail: llrw@umn.edu (LLRW)}
Jori Liesenborgs$^{2}$
\\
$^{1}$School of Physics and Astronomy, University of Minnesota, 116 Church Street, Minneapolis, MN 55455, USA \\
$^{2}$UHasselt - tUL, Expertisecentrum voor Digitale Media, Wetenschapspark 2, B-3590, Diepenbeek, Belgium\\
}

\date{Accepted XXX. Received YYY; in original form ZZZ}
\pubyear{2018}
\begin{document}
\label{firstpage}
\pagerange{\pageref{firstpage}--\pageref{lastpage}}
\maketitle

\begin{abstract}
Multiple image gravitational lensing systems with measured time delays provide a promising one-step method for determining $H_0$. MACS J1149, which lenses SN Refsdal into a quad S1-S4, and two other widely separated images, SX and SY, is a perfect candidate. If time delays are pinned down, the remaining uncertainty arises from the mass distribution in the lens. In MACS J1149, the mass in the relevant lens plane region can be constrained by (i) many multiple images, (ii) the mass of the galaxy splitting S1-S4 (which, we show, is correlated with $H_0$), (iii) magnification of SX (also correlated with $H_0$), and  (iv) prior assumptions on the mass distribution. Our goal is not to estimate $H_0$, but to understand its error budget, i.e., estimate uncertainties associated with each of these constraints. Using multiple image positions alone, yields very large uncertainty, despite the fact that the position of SX is recovered to within $\!\le\!0.036$\as~(rms $\!\le\!0.36$\as) by \grale~lens inversion. Fixing the mass of the galaxy that splits S1-S4 reduces $1\sigma$ uncertainties to $\sim 23\%$, while fixing the magnification of SX yields $1\sigma$ uncertainties of $32\%$. We conclude that smaller uncertainties, of order few percent, are a consequence of imposing prior assumptions on the shapes of the galaxy and cluster mass distributions, which may or may not apply in a highly non-equilibrium environment of a merging cluster. We propose that if a measurement of $H_0$ is to be considered reliable, it must be supported by a wide range of lens inversion methods.
\end{abstract}


\section{Introduction}   

Multiple image gravitationally lensed systems can be used to measure $H_0$ in one step \citep{ref64}, and completely independently of the distance ladder. The two inputs are the observed time delays between multiple images, and the mass distribution in the lens. In galaxy-size lenses, with quasars as sources, time delay measurements are currently uncertain at the few\%--10\% level \citep{bon16,bon18,cor18}. 

In galaxy clusters, with supernovae as sources, the uncertainty in time delay measurement can be reduced because clusters are large, and hence all scales, including time scales, get blown up, reducing the fractional errors in time delay measurements.

The first supernova with spatially resolved multiple images \citep{kelly15}, nicknamed Refsdal, was discovered in Grism Lens Amplified Survey from Space (GLASS; PI T. Treu) as four quad images, S1-S4, surrounding an elliptical galaxy, in the Hubble Frontier Field (HFF; PI J. Lotz) galaxy cluster MACS J1149.5+2223 (hereafter MACS J1149). Refsdal was later confirmed to be a supernova, of a type similar to that of SN 1987A \citep{kelly16b}. The time delays between all the images of the quad were measured to be a few days to 3-4 weeks \citep{rod16}. Viewed on cluster scale, images S1-S4 form in the second lowest minimum of the arrival time surface. Based on mass models \citep{sha15,die16,gri16,jau16,ogu15,kaw16}, a cluster-scale saddle-point image, SX, was predicted \citep{treu16}, and then observed \citep{kelly16a}, at the time and position consistent with predictions. These successes prompted early estimates of $H_0$ from Refsdal \citep{veg18,grillo18}. The same cluster also hosts a transient, arising from a highly magnified, macro- and micro-lensed, massive high-redshift star \citep{kelly18}, whose images, Icarus (LS1/Lev16A) and Iapyx (Lev16B), separated by $\sim 0.3$\as, straddle a nearby portion of the cluster critical line. The supernova and the transient source live in the same, triply imaged, face-on spiral galaxy at $z=1.489$. The whole system of lensed images in MACS J1149 is spectacular and unique, and needs as much detailed investigation as possible.

This host galaxy cluster at $z = 0.542$ was first identified as one of 12 most distant X-ray luminous clusters detected at $z>0.5$ by the Massive Cluster Survey, MACS \citep{ebe07}. It is a double merger \citep{gol16}, and one of the most complex merging clusters known \citep{ogr16}. It was observed with the Advanced Camera for Surveys on HST, and analyzed and modeled by \cite{zit09}, \cite{smi09} and \cite{rau14}. They found three images of an extended, and surprisingly undistorted background spiral at $z=1.489$, which is also the host of both SN Refsdal and the transient. 

The time delay between Refsdal images S1 and SX can be determined to a 1-2\% precision (Kelly et al., in preparation). However, the mass distribution is an additional, and largest source of uncertainty. (The global cosmological model is an additional, but smaller source of uncertainty.) Several mass models reproduce the lensed images within observational uncertainties, but lensing degeneracies can prevent an accurate and precise determination of $H_0$.  Models that quote low uncertainties usually rely on lens inversion methods that break many degeneracies through the use of parametric mass distribution assumptions.

Instead of estimating the value of $H_0$, the goal of this paper is to investigate how much precision in $H_0$ is possible based solely on the lensed image positions, and relations implied by lensing reconstructions, and how much precision is brought about through the use of priors, i.e., the assumptions about the density profiles of cluster galaxies and cluster dark matter. We do this by comparing the precision achieved by free-form \grale~reconstructions, which are based on image positions only, and the precision yielded by parametric mass models, which use lensed images as well as priors on mass distribution. To isolate the constraining power of lensed images on the mass distribution, we assume that the parameters that would otherwise give rise to additional uncertainty---namely, the cosmological model and measured time delays---are fixed. An understanding of the error budget is important when striving at a precision level of a few percent in $H_0$.

Ideally, to investigate the constraints on $H_0$ brought about by the lensed images, one would like to use a technique which is free of any subjective assumptions. Since this is not possible, we use a method which does not make strong prior assumptions, \grale. Because we use a specific method, the uncertainties probed in this paper do not necessarily represent the general case for the uncertainties on $H_0$. However, of all existing methods, \grale~is probably best positioned to address the question of uncertainties because its internal uncertainties come closest to spanning the range of uncertainties of all other methods combined. This was first noticed in \cite{rod15}, where \grale's uncertainties on the magnification of SN Ia HFF14Tom encompassed those of all other methods, as well as the observed value (see their Figure 6). This study, in somewhat modified form, was later extended to a grid of lens plane locations in two HFF clusters, by \cite{pri17}. Their Figures 14 and 15 show that of all the lens models (6-7 in total), \grale's uncertanties are by far the most consistent with systematic uncertainties.

We adopt the concordance $\Lambda$CDM cosmological model: flat, matter density, $\Omega_m=0.3$, cosmological constant density, $\Omega_\Lambda=0.7$, and the dimensionless Hubble constant $h=0.7$, such that $H_0=100h~$km~s$^{-1}$~Mpc$^{-1}$. At the redshift of the cluster, $z_l=0.542$, 1\as~translates into 6.36 kpc. The source galaxy of Refsdal is at $z_l=1.489$, and the critical surface mass density for sources at that redshift is $\Sigma_{\rm crit}=[c^2/(4\pi G)] D_{\rm os}/(D_{\rm ol}D_{\rm ls})=0.496$ g~cm$^{-2}$, where $D$'s are the angular diameter distances between the observer, $o$, lens, $l$, and source, $s$. We express the projected surface mass density in the lens in units of $\Sigma_{\rm crit}$, and denote that dimensionless quantity by $\kappa$.

\begin{figure*}    %% figure 1
\centering
\vspace{-5pt}
\includegraphics[width=0.99\linewidth]{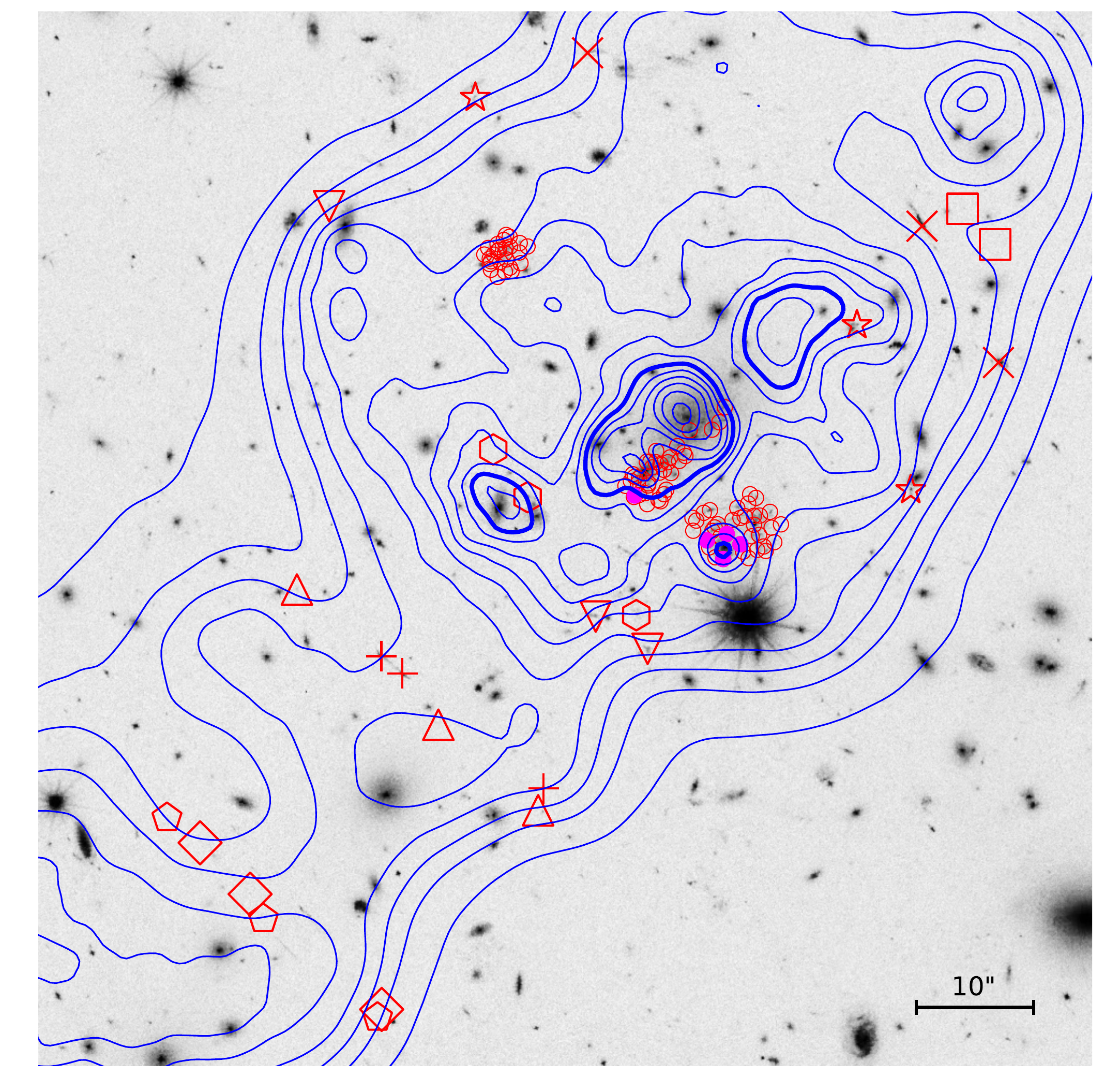}
%\vspace{-15pt}
\caption{Multiple images used in this work (red symbols) are superimposed on an HST image of MACS J1149. Same symbol types identify images belonging to the same source. Three sets of about 33 empty circles are the knots of the spiral galaxy at $z_s=1.489$. Magenta solid dots show the quad S1-S4 and SX image of supernova Refsdal. The two sources with photometric redshifts produce the 6 images in the lower left corner of the plot. The large diamonds correspond to the images of the $z=4.8$ source, while the pentagons are those of the $z=6.5$ source. The projected isodensity contours of the \hres~maps, at intervals of $\delta\kappa=0.1$ (for $z_s=1.489$), are shown in blue, with the thick contour representing $\kappa=1$. The 10\as length scale shown in the figure corresponds to 63.6~kpc at the redshift of the cluster, $z_l=0.542$.}
\label{images}
\end{figure*} 

\section{Grale-based inversions}

\subsection{Method \& input}

\grale~is a flexible, free-form, adaptive grid lens inversion method, based on a genetic algorithm. The code is publicly available, and open source. It has been described extensively in previous works \citep{lie06,lie07,moh14,men17}. Here, we give a brief summary. A \grale~run starts with an initial coarse uniform grid in the lens plane populated with a basis set, such as projected Plummer density spheres {\citep{plu1911}. Each grid cell has a single Plummer sphere, with the size matching the cell size. Plummer spheres are chosen because they have constant central density, and rapidly, but smoothly falling off density at larger radii.} As the code runs, the denser regions are resolved with a finer grid, with each cell given a Plummer with a proportional width. The initial trial solution, as well as all later evolved solutions are evaluated for genetic fitness, and the fit ones are cloned, combined and mutated. The final map consists of a superposition of many Plummers, typically several hundred to a couple of thousand, each with its own size and weight, determined by the genetic algorithm. Note that there is no one-to-one correspondence between Plummer spheres and astrophysical mass concentrations, like cluster-wide dark matter distributions, or individual galaxies. Plummer spheres are building blocks of these structures. We note that \grale~does not use any regularization.

For this paper, \grale~uses a combination of two types of fitness measures: based on (i) image positions, and (ii) the null space. (i) Images are assumed to be point-like in this version of \grale, and to assign this fitness measure the observed point images are first projected back onto their source planes. If the points of the same sources lie closer together, the trial solution is considered to have a better fitness value. The distances between the points are not measured on an absolute scale, however. Instead, the size of the area of all backprojected points (of all sources) is used as a length scale. This helps avoid mass-sheet degenerate solutions (see below) having the advantage by introducing a relatively large mass-sheet component, as this would project all image points onto a smaller area. (ii) A trial solution under consideration may perform well regarding fitness measure (i), but still predict unobserved additional images. To take this into account, an area that is larger than the region of the observed images is subdivided into a grid of triangles. By counting the number of backprojected triangles that overlap with the envelope of the backprojected image points, trial solutions can be penalized if they predicted additional images. \grale~uses a so-called multi-objective genetic algorithm \citep[see e.g.,][]{lie07} to optimize these two fitness measures at the same time.

\begin{figure*}    %%   figure 2
\centering
\vspace{-5pt}
\includegraphics[width=0.49\linewidth]{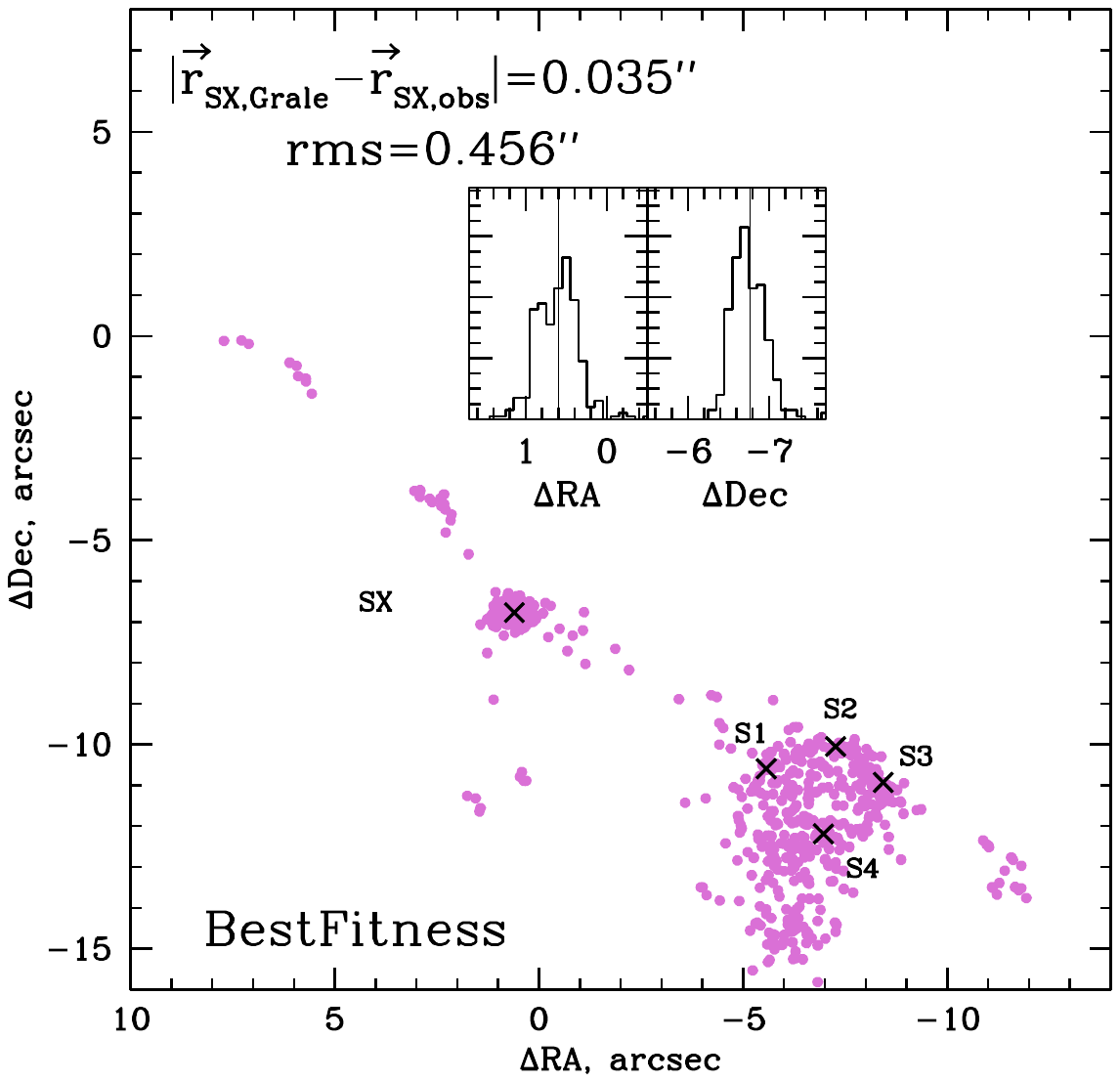}
\includegraphics[width=0.49\linewidth]{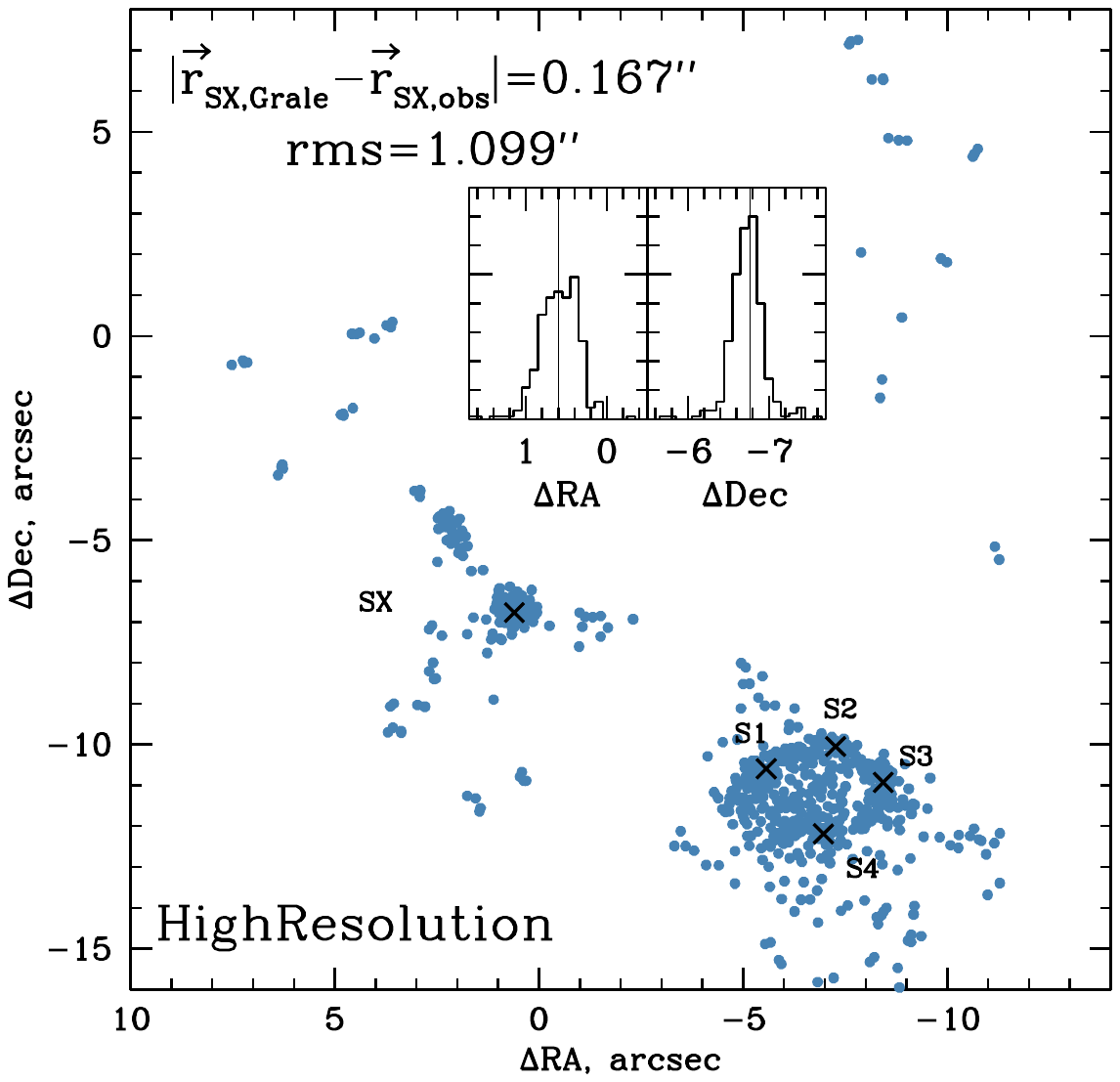}
\vspace{-55pt}
\caption{{\it Left:} The points are all the images of Refsdal predicted by the \bfit~set of \grale~mass maps from 75 independent runs. To obtain these, each of the observed S1-S4 images was traced back to the source plane, and used individually as sources to generate predicted images. The total number of images is over 1000. The observed SX, S1-S4 are marked with black crosses. See Section~\ref{predic} for a description of how the average position and the rms of the predicted SX (displayed in the figure) were calculated. The insets show the distribution of the $\Delta$RA and $\Delta$Dec of predicted SX images. Vertical lines mark the observed values. {\it Right:} Same as the left panel, but for the \hres~set of \grale~maps from the same 75 runs. 
}
\label{allims}
\end{figure*} 

\begin{figure*}    %%   figure 3
\centering
\vspace{-5pt}
\includegraphics[width=0.49\linewidth]{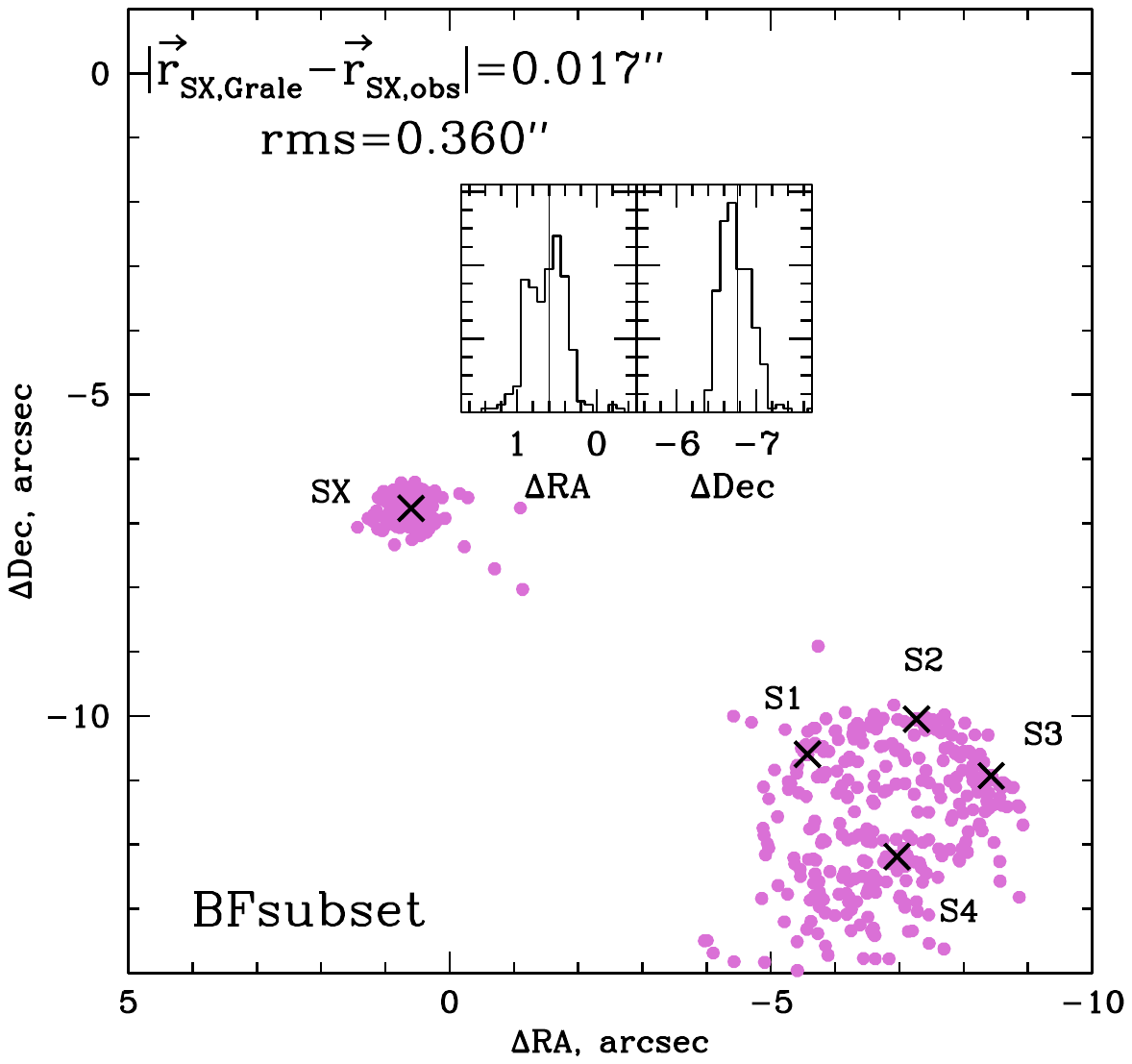}
\includegraphics[width=0.49\linewidth]{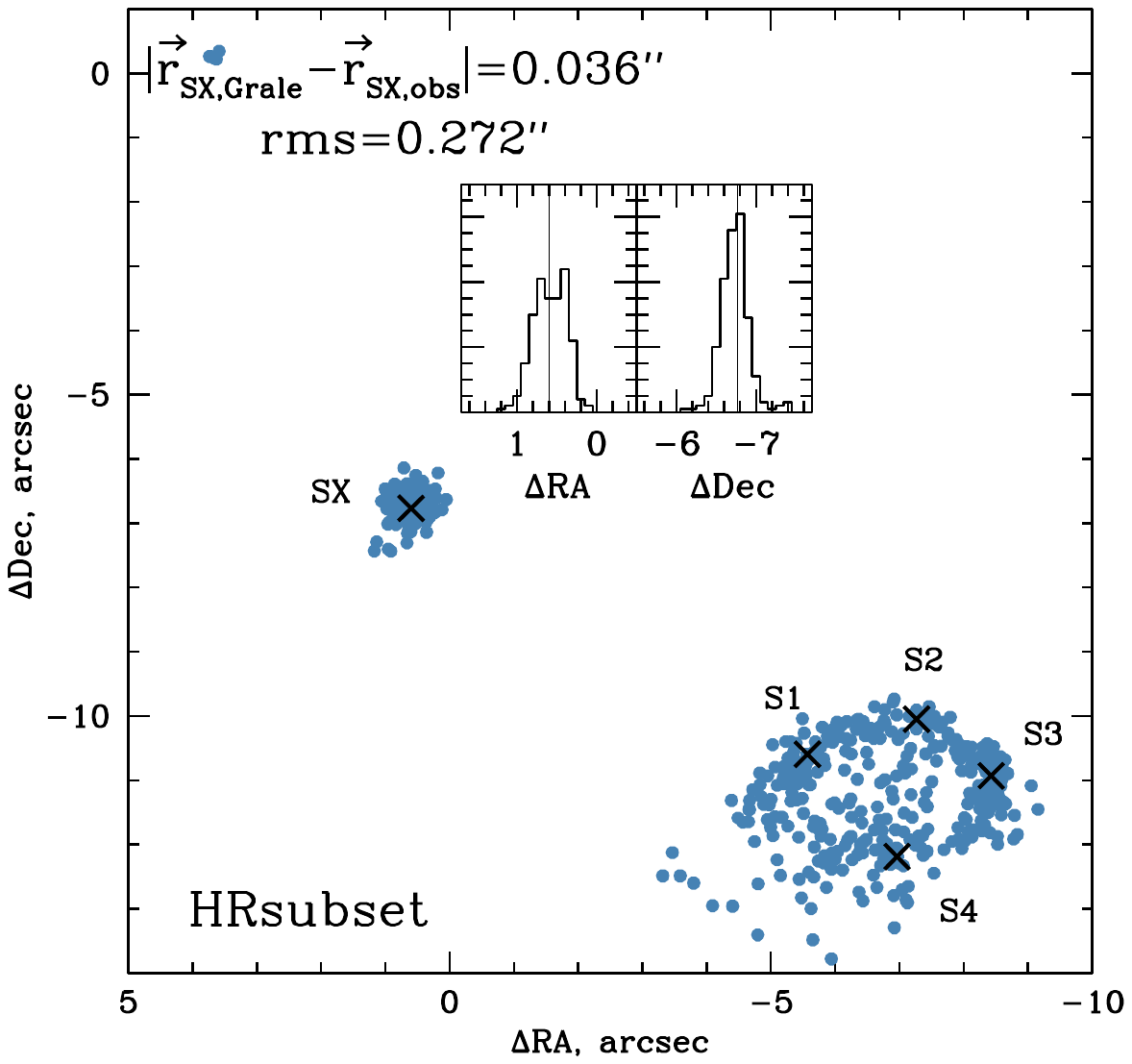}
\vspace{-60pt}
\caption{Of the 75 runs, we selected those where all 4 predicted SX images (one from each of S1-S4) had negative parity (i.e., were saddle points in the arrival time of the cluster-wide potential), and had no other predicted images within $5$\as of observed SX. This selection leaves 67 \bfit~({\it left}) and 59 \hres~({\it right}) maps. The displacement between the observed and predicted SX, and rms of the scatter of predicted positions are shown in the figure. The insets in both panels show the distribution of the $\Delta$RA and $\Delta$Dec of predicted SX images in these maps. Vertical lines mark the observed values.}
\label{selectedims}
\end{figure*} 

We used the following as input to our models: positions of all 93 multiple images of knots in the spiral source galaxy from Table 4 of \cite{treu16}, and positions of the 4 Refsdal S1-S4 images. All these are at $z=1.489$. Estimating source redshifts using lens inversion techniques can lead to biased estimates, because of lensing degeneracies \citep{wil17}. To avoid that we used 20 images that had spectroscopic redshifts \citep[also from][]{treu16}. For sources at very high redshifts, the uncertainties in redshifts become less important; the fractional change in $\Sigma_{\rm crit}$ between $z_s=2$ and $4$ is 0.164, while between $z_s=4.5$ and $6.5$ it is 0.041, i.e., four times smaller. To supplement our image set, we included two sources (6 images) with photometric redshifts of 4.8 and 6.5, respectively. In all, we used 123 images from 40 sources. These image data presented in \cite{treu16} were based on spectroscopic redshifts from VLT-MUSE \citep{gri16}, Keck DEIMOS and HST-WFC3 \citep{bra16,wang17,treu15}. The images we used are shown in Figure~\ref{images}. Their coordinates and redshifts are available in an online table accompanying this paper (the first 26 entries are images are from 9 sources with spectroscopic and 2 with photometric redshifts, the next 93 entries are the knots in the $z=1.489$ spiral galaxy, and the last 4 entries are the S1-S4 Refsdal images.)

\cite{treu16} and \cite{kelly16a} present results of several lens inversion models, predicting, in advance of the event, the location and arrival time of Refsdal image at SX, which was first detected on December 11, 2015. In order to directly compare our results to the published ones, we did not use the observed position of SX as a constraint. We also did not include the observed positions of the two transients, Icarus and Iapyx, as these had not yet appeared at the time of the published predictions. As all previous \grale~reconstructions, and in contrast to all but one other existing technique \citep{qui18}, our models do not use any information about galaxies in the lens or along the line of sight.

\subsection{Inversion results}

Each \grale~run consists of 9 sequential solutions, where the number of Plummers, and hence mass resolution, increases with each subsequent solution. We carry out a total of 75 of these runs\footnote{These reconstructions are not the same as the v.4 set submitted to HST in early 2017. The v.4 reconstructions did not include S1-S4, and had a somewhat different, and smaller input image set.}, each of which is started with its own random seed, so the runs are independent. Some runs slightly vary the number of Plummers, as well as the center and size of the reconstruction grid.

In each of the 75 runs we identify two solutions and the corresponding mass maps. (1) The first is the solution that has the best fitness of all the 9 sequential solutions. For MACS J1149 runs, these are usually the 4th, 5th or 6th solution in a specific run. We call these the \bfit~reconstructions, or maps. (2) The second solution is identified from the two last solutions---8th and 9th---as the one with the better fitness.

We call these the \hres~maps, because latter solutions have finer grids and more Plummers making up the total mass map, which implies higher mass resolution. Higher resolution will be advantageous in MACS J1149 because it hosts many closely spaced images in the vicinity of Refsdal's quad S1-S4. Selections (1) and (2) yield 75 maps each. (In 6 of the 75 runs, the \bfit~and the \hres~maps are the same, so the total number of unique maps is 144, instead of 150. Because this is a small fraction of the total, we do not eliminate the redundant maps.) We work with two sets of maps instead of just one in order to obtain more encompassing uncertainties, closer resembling systematic uncertainties.

From these two sets of maps, \bfit~and \hres, we construct one subset each, BFsubset and HRsubset, respectively. We select these based on how well they reproduce the topology of the cluster-wide arrival time surface for Refsdal (see Section~\ref{predic} for details). After briefly presenting the results of the full \bfit~and \hres~of maps, the rest of the analysis in this paper concentrates on BFsubset and HRsubset only.

\section{Image plane predictions}\label{predic}

The left and right panels of Figure~\ref{allims} show the distribution of all \grale~predicted images of Refsdal for the \bfit~and \hres~reconstructions, respectively. (Images corresponding to SY are outside the box shown, and are not considered in this paper, because they were not observed.) These images were obtained as follows. For each of the \grale~reconstructions, each of the four observed S1-S4 images was traced back to the source plane, and used as a source to generate lens plane images. Note that not all of these sources produced 4 images of the S1-S4 quad, but all produced at least one image very close to observed SX. 

As expected, most of the images are associated with the S1-S4 quad, and the location of SX. The images of the S1-S4 quad are not reproduced very well, most likely because \grale~does not have enough spatial resolution, even in \hres~maps. The total number of images in each of the two panels is over one thousand. (In principle, each of the backprojected images of the quad should generate 5 images, and the SX image, for a total of 1800 images.) There are also a handful of images far from S1-S4 and SX; these are most likely spurious, i.e. have no observed counterpart. Also as expected, \hres~has more spurious images than \bfit, because the former maps are allowed to have more small scale density fluctuations.

\begin{table}
\centering
\begin{tabular}{|l|c|c|}
        \hline {\bf Model}         & $\langle|\vec r_{\rm{SX,obs}}-\vec r_{\rm{SX.Grale}}|\rangle$ & rms\\
        \hline \bfit~ - all 75 maps        & 0.035\as   & 0.456\as \\
        \hline BFsubset - subset of 67 maps   & 0.017\as   & 0.360\as \\
        \hline \hres~ - all 75 maps     & 0.167\as   & 1.099\as \\
        \hline HRsubset - subset of 59 maps & 0.036\as   & 0.272\as \\
\end{tabular}
\caption{Summary of the four sets of reconstructions discussed in the paper. For each set we quote how well it reproduces the position of Refsdal image SX, which was not part of \grale~input. The selection of the maps in the second and fourth row is described in Section~\ref{predic}.}
\label{tab1}
\end{table}

The right panel of Figure~\ref{allims} has a few predicted images in the upper right, scattered widely around $(-8,3)$. These could be the fourth image of the cluster-wide Refsdal quad, with the fifth image hidden in the BCG. However, only a small fraction of our reconstructions have these images. We conclude that it is most likely that SY, S1-S4, and SX comprise a naked cusp, triple-image system split by the cluster. There are no additional images of Refsdal due to the cluster-wide potential. 

The concentration of predicted images near the location of observed SX is very high, implying that the predicted position of SX is very well localized. To calculate the average predicted position of SX we first need to identify these images. For every \grale~map, and for every observed S1-S4, which we backproject to the source plane and re-lens, we found the \grale~predicted image closest to observed SX. These 4 images per map are our predicted SX locations. Note that using the observed SX in this way does not bias the estimation of \grale~predicted position. The two insets in each of the two panels of Figure~\ref{allims} show the histograms of $\Delta$RA and $\Delta$Dec of predicted SX images. The position of observed SX is indicated by the vertical lines. The distance and rms of the observed and \grale~predicted SX images are displayed in the figure, as well as in Table~\ref{tab1}.

Starting from the full set of 75 \bfit~and 75 \hres~maps, we select those where all 4 predicted SX images have negative parity (i.e., are saddle points in the arrival time of the cluster-wide potential), and have no other predicted Refsdal images within $5$\as~of observed SX. This refinement of the full sample leaves 67 BFsubset and 59 HRsubset maps. All subsequent analyses in the paper are done using these two subsets.

Figure~\ref{selectedims} is similar to Figure~\ref{allims}, but for the BFsubset and HRsubset. Even though no cuts other than the two discussed above were imposed, BFsubset reconstructions have no spurious images. The only remaining extra spurious images are the 4 images in the HRsubset (right panel of Figure~\ref{selectedims}), around $(3.7,\,0.3)$. All 4 arise from a single map, out of a total of 126, and therefore will have minimal effect on the analysis in this paper.

The displacement between observed and predicted SX in our BFsubset and HRsubset ($\le 0.036$\as; see Table~\ref{tab1}) appear to be smaller than in other existing models ($\sim 0.1-0.3$\as); the smallest published offset is that of the \cite{gri16} model \citep[see][]{kelly16a}, and is $\sim 0.06$\as. Our rms values are in the range $\le 0.36$\as, while other models have rms values between $0.26$\as~and $0.9$\as \citep{kelly16a}.

\begin{figure}    %%   figure 4
\centering
\vspace{-5pt}
\includegraphics[width=0.99\linewidth]{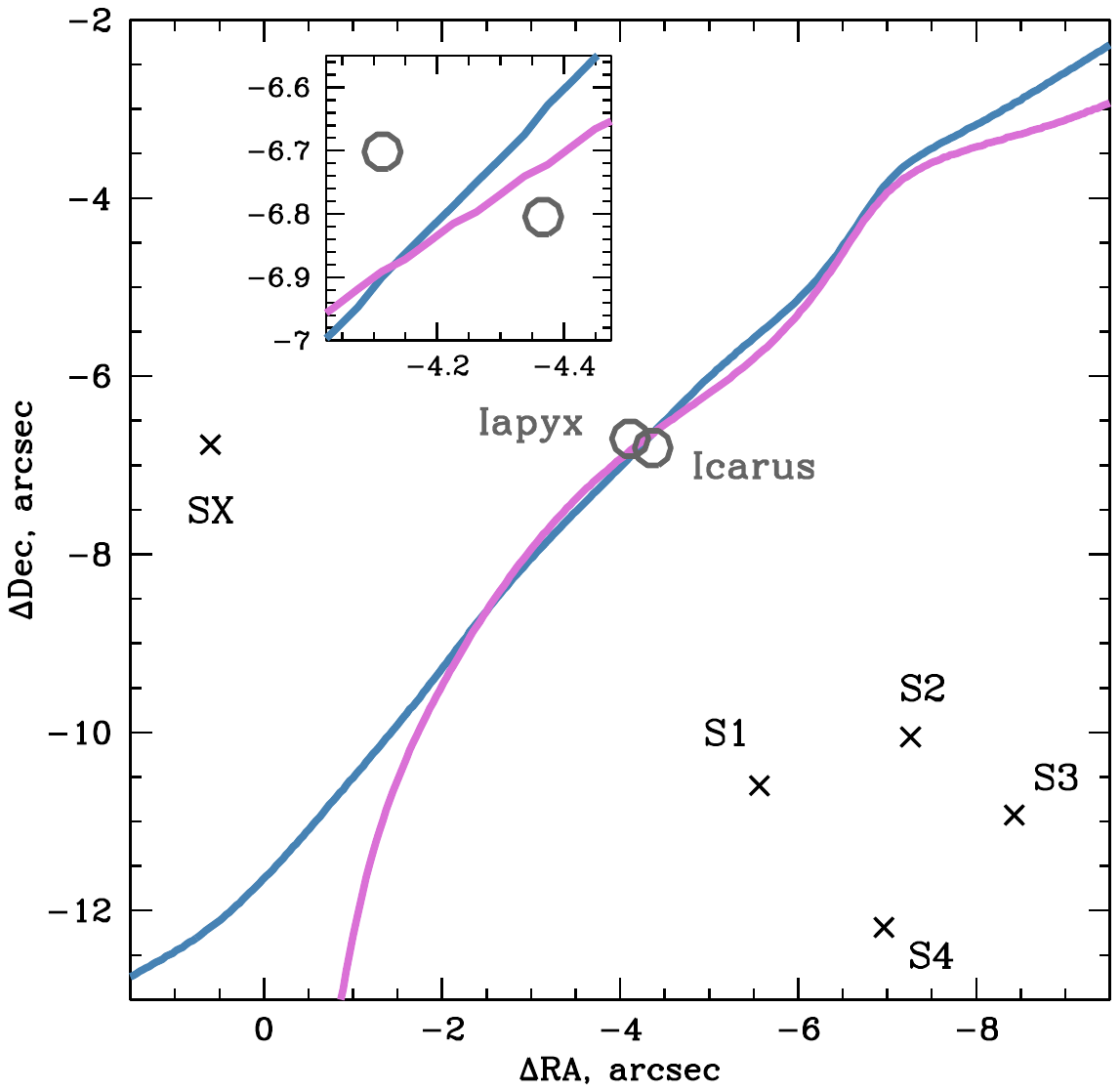}
\vspace{-65pt}
\caption{Cluster critical curve in the vicinity of Refsdal images, which are marked with black crosses. Gray empty circles are the Icarus and Iapyx locations, and were not used in \grale~reconstructions. (S1-S4 were used, but SX was not used.) The pink and blue curves are based on the 67 (59) maps of BFsubset and HRsubset reconstructions. The inset shows a zoomed in region around the transients.}
\label{cc}
\end{figure}

Before we leave the image plane, let us look at the critical curve, for Refsdal's $z_s$. Figure~\ref{cc} shows the critical curves for BFsubset (pink), and HRsubset (blue) curves, and a zoom in the inset. Observed positions of Icarus and Iapyx were not used as model inputs, and yet the \grale~critical curves go between them. Along the line connecting these transients, the separation between the 2 curves is $\sim 0.05$\as. By comparison, the separation between the four models presented in \cite{kelly18} \citep[namely,][]{jau16,kaw16,zit15,kee10} is $\sim 0.25$\as. Furthermore, \grale~critical curves pass between the two transients, implying that Icarus and Iapyx are counterimages of the same source. Two of the published models \citep{jau16,kaw16} make a similar prediction, while the other two models have both the transients to the North East of the critical curves.

\begin{figure}    %%   figure 5
\centering
\vspace{-5pt}
\includegraphics[width=1.0\linewidth]{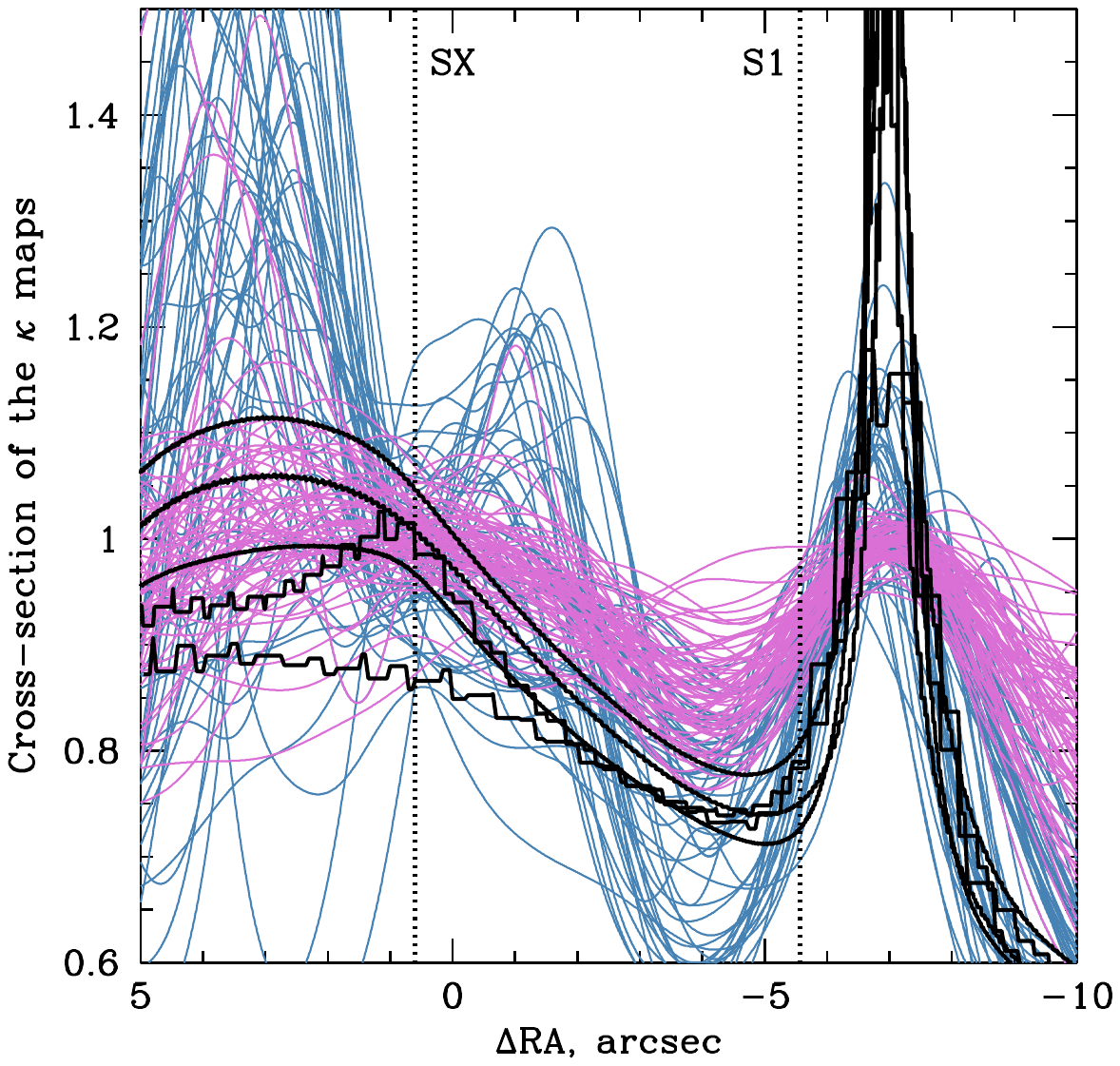}
\vspace{-65pt}
\caption{Cross-sections of the $\kappa$ density map taken along the line connecting \grale~predicted positions of SX, and S1. The line is parametrized by the $x$-coordinate, $\Delta$RA, from the center of the cluster. The two vertical black dotted lines show the positions of the observed SX and S1. The pink and blue curves show the 67 BFsubset and 59 HRsubset reconstructions, respectively. The black curves show all five of the latest (v3 and v4) publicly available ({\tt https://archive.stsci.edu/prepds/frontier/lensmodels/}) reconstructions; from top to bottom (looking at the left edge of the plot) the models are: Sharon et al., Keeton et al., Glafic team, CATS team, and Diego et al. On the right side of the plot most of these models are indistinguishable.}
\label{denxparam}
\end{figure}

\section{grale time delay predictions}

In the previous section we saw that using only the image positions \grale~is able to predict the position of SX and the locus of the critical curve in the vicinity of the transient, accurately and precisely. In this section we examine how well image positions can constrain $H_0$, assuming the observed time delay between Refsdal's two images SX and S1 is known exactly, and the global cosmology is fixed.

\subsection{Generalized mass sheet and monopole degeneracies}

It is well known that the mass sheet degeneracy \citep[MSD;][]{fal85}, also known as the steepness degeneracy \citep{saha00}, is one of the lensing degeneracies that needs to be broken to measure $H_0$. Two mass maps, $\kappa_a(\vec r)$ and $\kappa_b(\vec r)$, are related by an MSD transformation if there exists a constant $\lambda$, such that $\kappa_b=\lambda\kappa_a+(1-\lambda)$, or equivalently, $\Sigma_b=\lambda\Sigma_a+(1-\lambda)\Sigma_{\rm crit}$ over the entire lens plane, where $\Sigma$ is the surface mass density of the lens in physical units. Such a transformation would leave most lens observables---image positions, flux ratios, time delay ratios---unchanged, but will rescale all time delays by $\lambda$, thereby scaling the estimated $H_0$ by $\lambda$.  Because $\kappa$ depends on the source redshift, the presence of multiply imaged sources at different redshifts breaks the degeneracy, as no single $\lambda$ works for all sources.

It was demonstrated in \cite{lie08a} and \cite{lie12} that a generalized version of the mass sheet degeneracy, gMSD, can be present despite sources at multiple redshifts. For gMSD, the added mass sheet is not simply $(1-\lambda)\Sigma_{\rm crit}$. Instead, it is a non-uniform disk, $\Sigma_{\rm gen}$, centered on $\vec\theta_c$, with its density profile arranged so that images at different redshifts and different locations $\vec\theta$ see the same total enclosed mass as they would with the regular MSD: $\Sigma_b(\vec\theta)=\lambda\Sigma_a(\vec\theta)+(1-\lambda)\Sigma_{\rm gen}(|\vec\theta-\vec\theta_c|)$. In the regions of the lens plane that have images of different redshifts located close to each other, gMSD is suppressed, but not eliminated. Its approximate version, agMSD, that reproduces the image properties not exactly, but within observational uncertainty, presents an even wider range of possible mass models. 

MSD and its kin, agMSD, are not the only degeneracies. In this paper we will also focus on the monopole degeneracy, MpD \citep{saha00,lie08b,lie12}. In regions of the lens plane with no images, the projected mass distribution can be reshaped in any circularly symmetric fashion, as long as the net mass remains the same, and no extra images are produced. Since any number of such operations, with different centers, radii, and shapes can be performed and superimposed, the monopole degeneracy is very powerful. 

An important difference between agMSD and MpD is that agMSD reshapes the lens mass distribution over the whole lens plane, while MpD redistributes mass between images only. The immediate consequence is that agMSD is not necessarily affected by the high density of images in the lens plane (as long as they are at approximately the same redshift), while MpD is suppressed in regions where the density of images is high. Other degeneracies can also be present, for example the source plane transformation \citep[SPT;][]{ss14}, which is related to MSD, and also reshapes the mass continuously across the lens plane. For brevity, we will talk about agMSD and MpD only, noting that other degeneracies are also possible.

\begin{figure*}    %%   figure 6
\centering
\vspace{-5pt}
\includegraphics[width=0.49\linewidth]{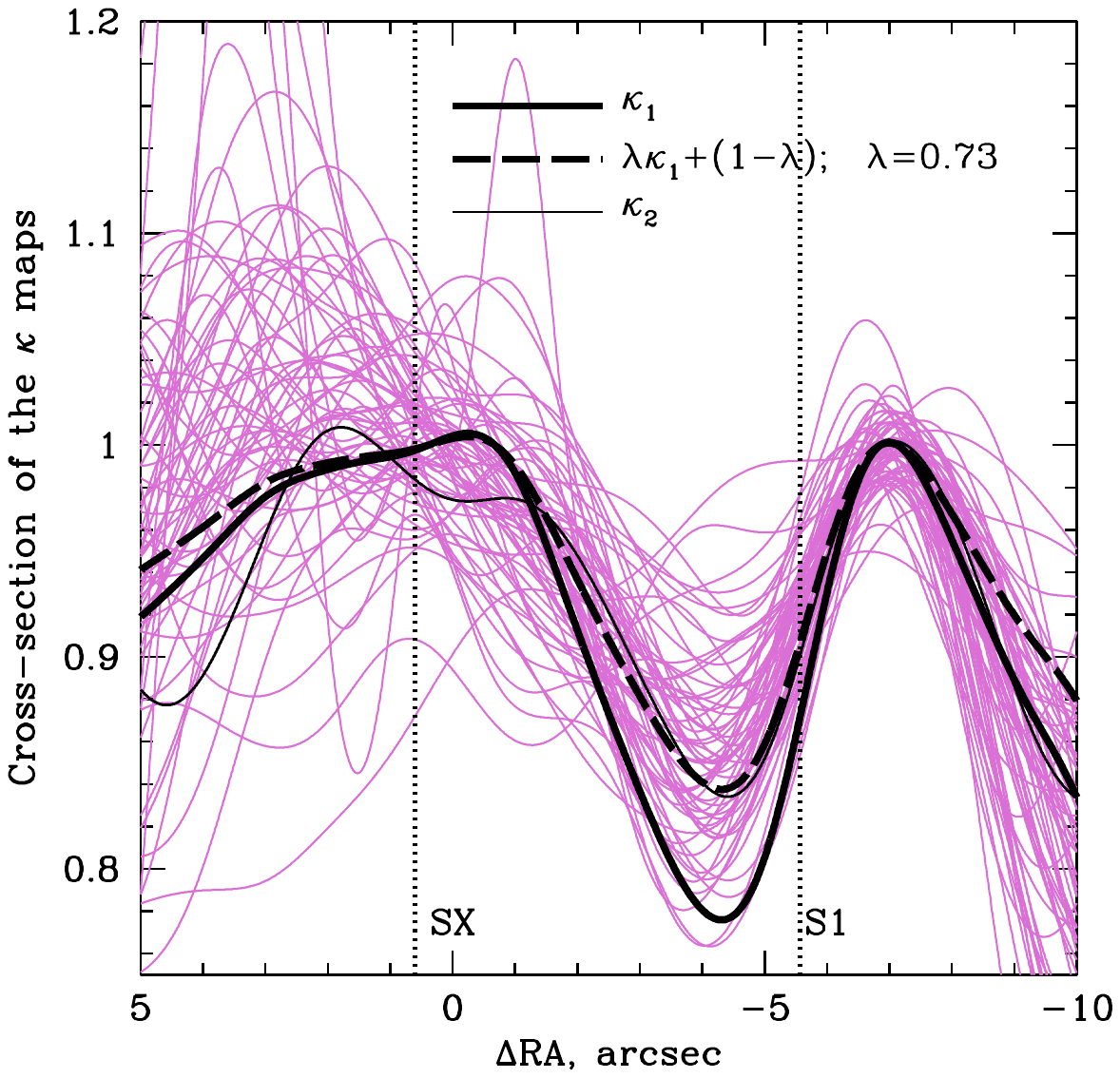}
\includegraphics[width=0.49\linewidth]{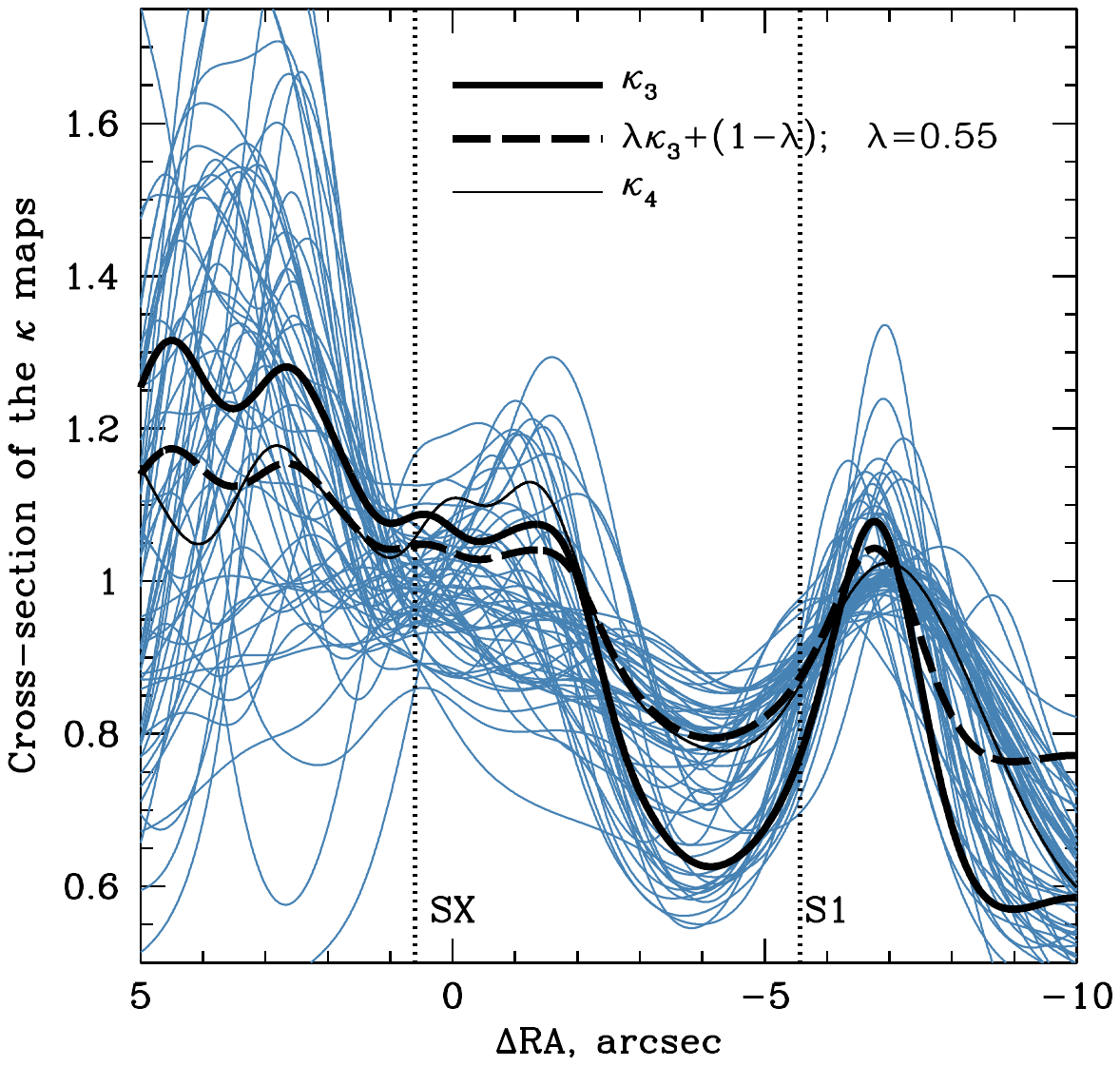}
\vspace{-50pt}
\caption{Similar to Figure~\ref{denxparam}.
{\it Left:} The pink curves show the 67 \grale~maps of the BFsubset. The thick solid black curve highlights one of these maps (which we call $\kappa_1$), while thin solid black curve highlights another map ($\kappa_2$). These two are related by approximate, generalized MSD, because $\lambda\kappa_1+(1-\lambda)$ curve, shown as the thick dashed black curve, is very similar to $\kappa_2$. {\it Right:} Same, but for the 59 HRsubset maps. Note that the extent of the vertical axes are different in the two panels.}
\label{denxsec}
\end{figure*} 

\begin{figure*}    %%   figure 7
\centering
\vspace{-5pt}
\includegraphics[width=0.49\linewidth]{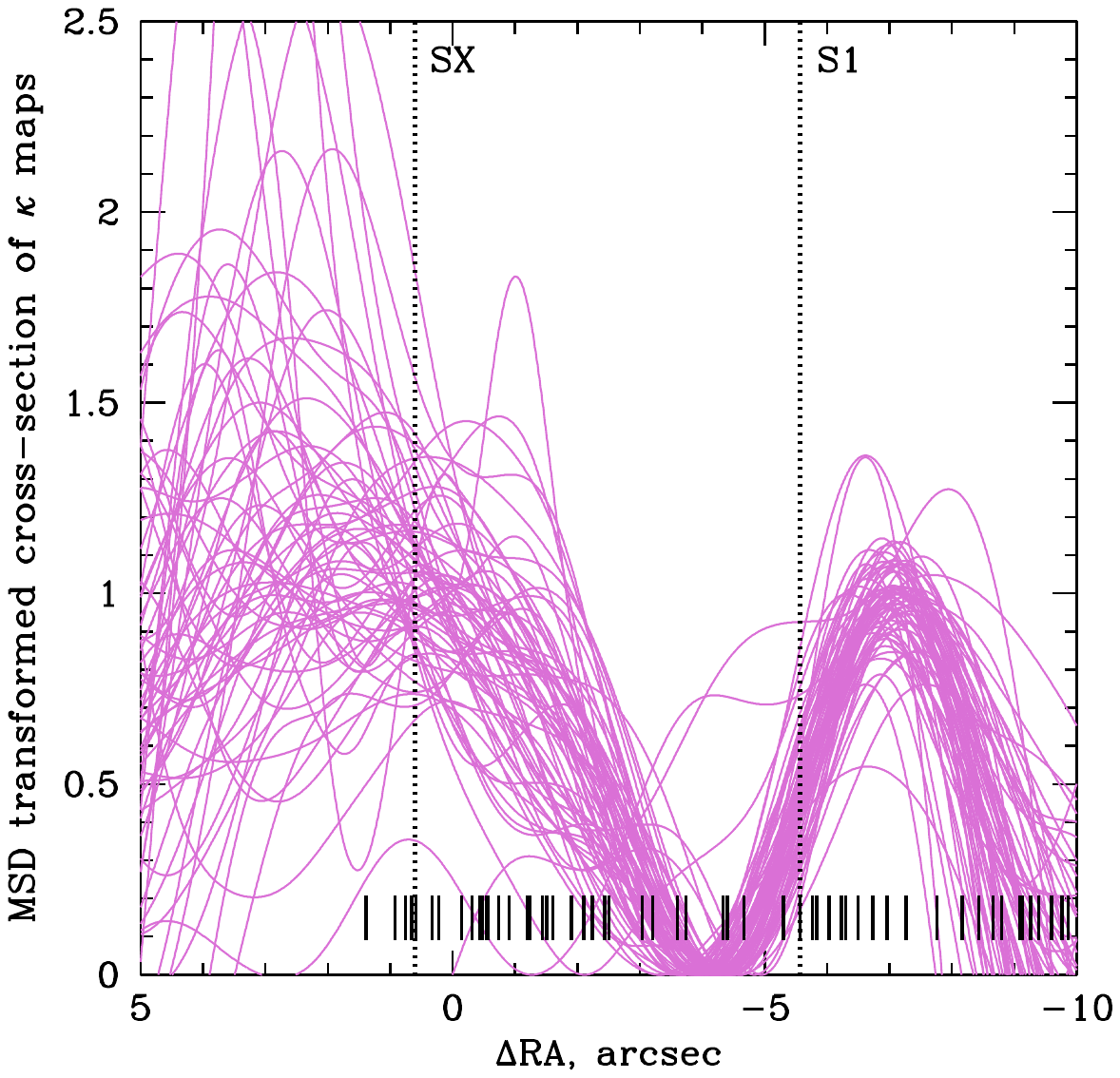}
\includegraphics[width=0.49\linewidth]{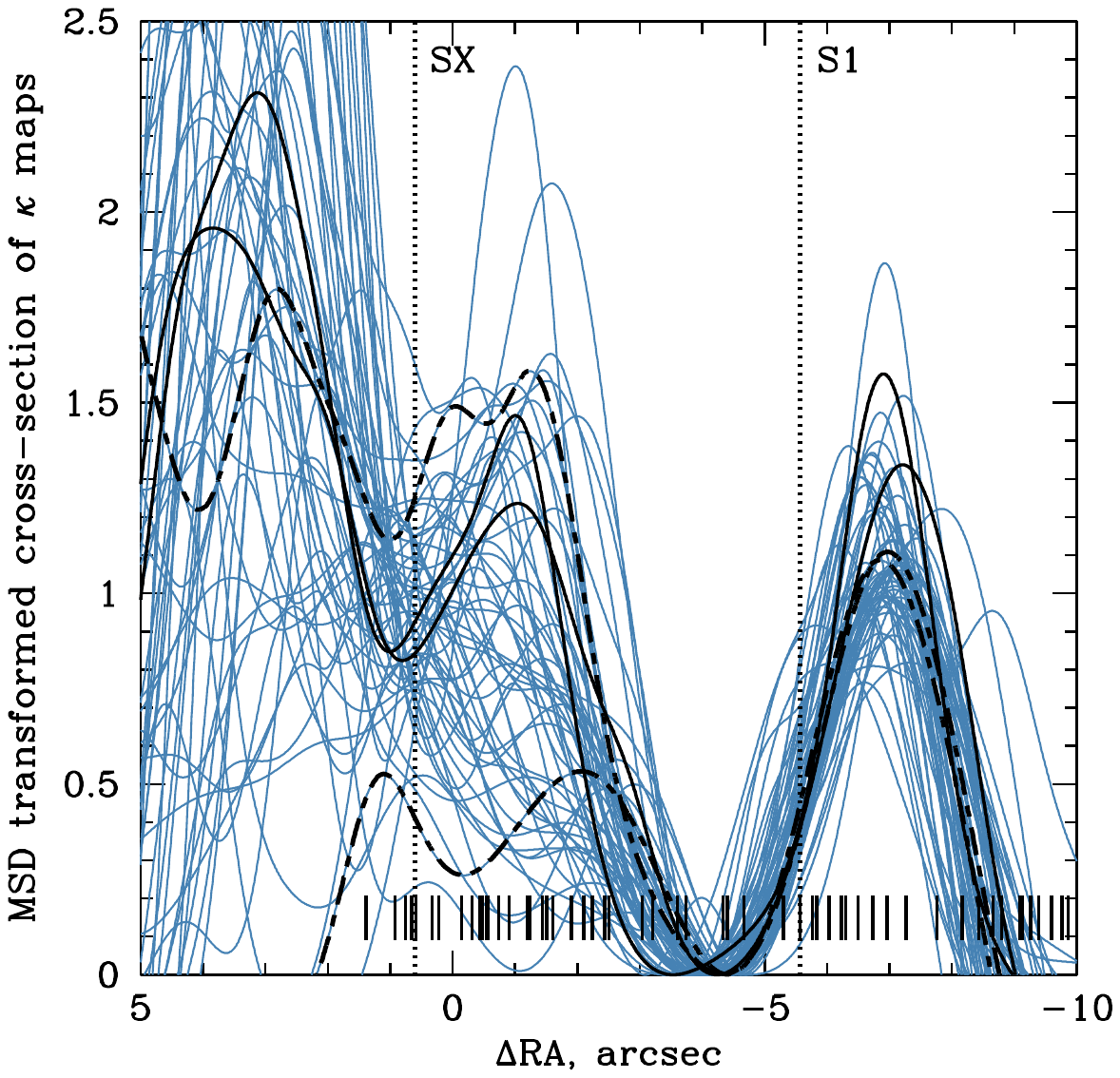}
\vspace{-50pt}
\caption{{\it Left:} Similar to Figure~\ref{denxsec}, but here we have taken out MSD, so that other degeneracies become visible. To do that, for every one of the 67 BFsubset $\kappa$ cross-sections presented in the left panel of Figure~\ref{denxsec}, we find $\lambda$ such that the curve $\lambda\kappa+(1-\lambda)$ attains $\kappa=0$ at some location within the $\Delta$RA range shown in the figure. The bar code at the bottom shows the locations of multiple images within $\pm 6$\as from the SX-S1 line, mapped on to the $\Delta$RA axis. {\it Right:} Same as the left panel, but for the 59 HRsubset maps. The four cross-sections highlighted in black are the reconstructions labeled A, B (solid lines), and C, D (dot-dashed lines) in Figures~\ref{kappadifftd} and \ref{massmap}.}
\label{MSDdenx}
\end{figure*} 

\begin{figure*}    %%   figure 8
\centering
\vspace{-5pt}
\includegraphics[width=0.49\linewidth]{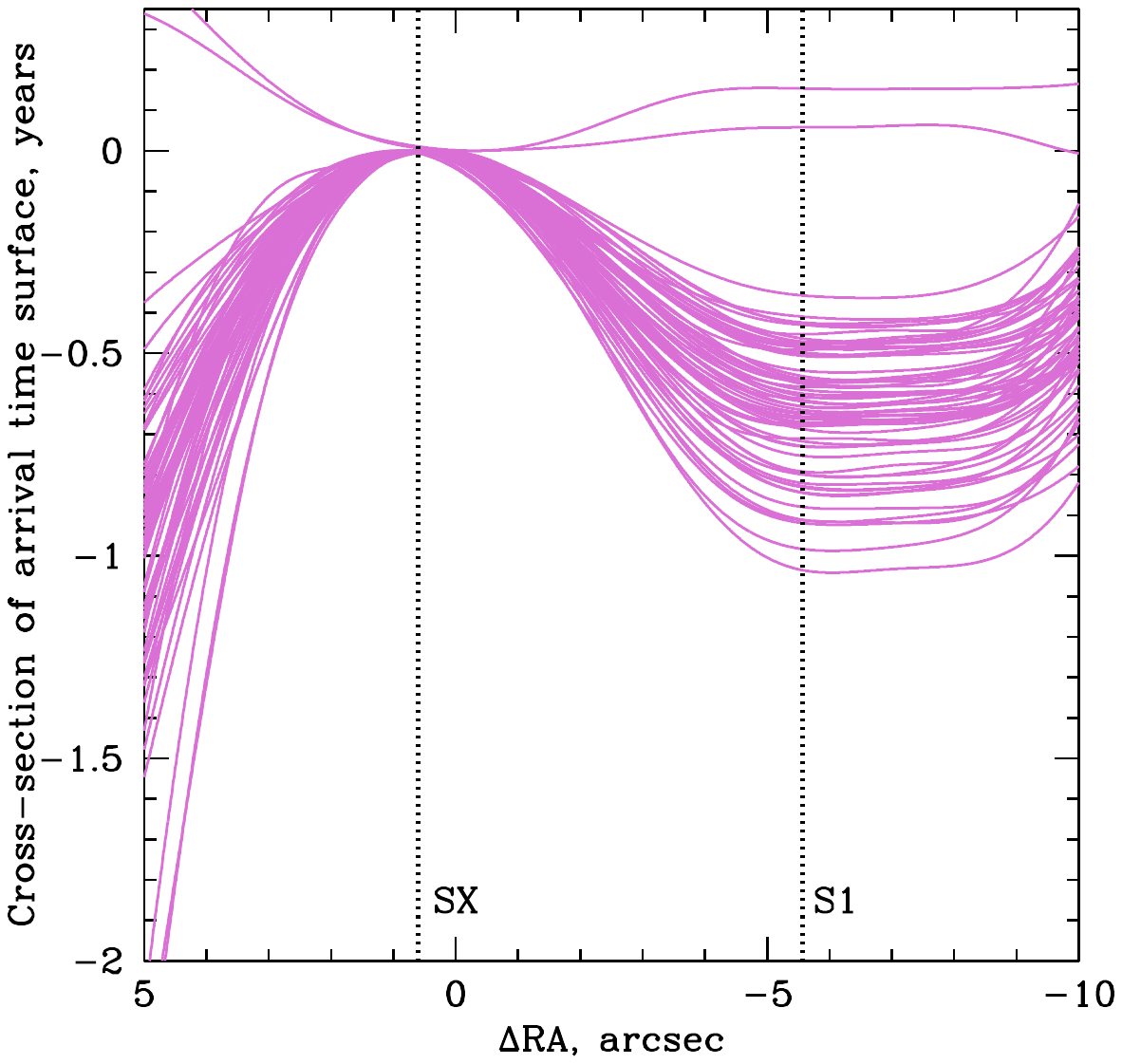}
\includegraphics[width=0.49\linewidth]{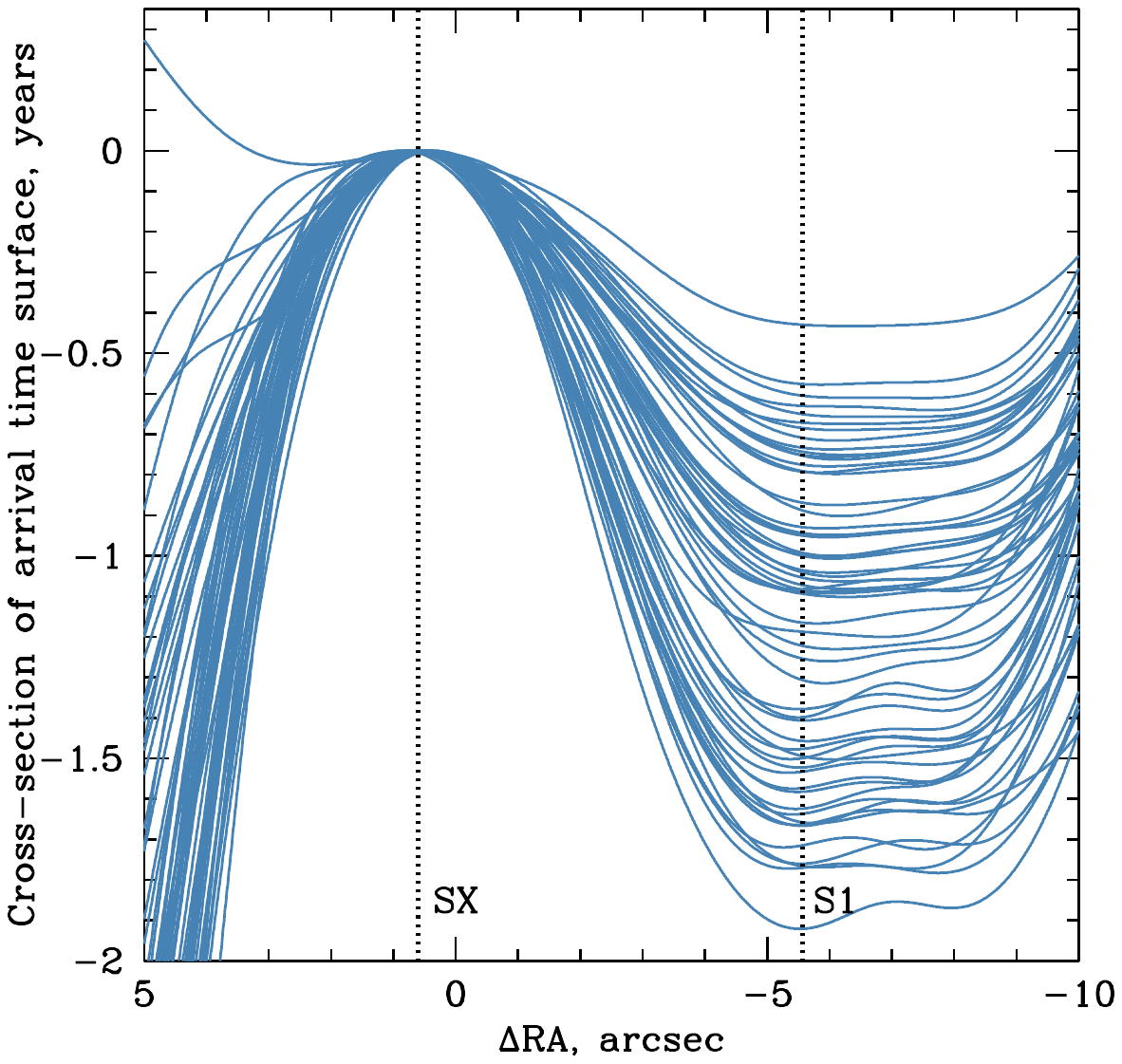}
\vspace{-50pt}
\caption{{\it Left:} Cross-section of the arrival time surface of the 67 LFsubset maps, taken along the line connecting \grale~predicted positions of SX, and S1. The line is parametrized by the $x$-coordinate, $\Delta$RA, from the center of the cluster. The two vertical black dotted lines show the positions of the observed SX and S1. {\it Right:} Same as the left panel, but for the 59 BFsubset maps. The scale is the same in both panels.}
\label{tdxsec}
\end{figure*} 

\subsection{Illustrating agMSD and MpD in grale reconstructions}\label{illus}

We will demonstrate the presence and effect of agMSD and MpD using our \grale~maps. The vertical axis of Figure~\ref{denxparam} shows the values of the $\kappa$ density maps taken along the line connecting \grale~predicted positions of SX, and S1. The line is parametrized by the $\Delta$RA coordinate, with respect to the center of the cluster. We call this and similar curves cross-sections, as they are one dimensional cuts through two dimensional surfaces, in this case, the $\kappa$ surface. The pink curves show the BFsubset maps, while the blue lines show the HRsubset.  As expected, all the \grale~curves follow the same general trend, for example, all show the presence of the galaxy at $\Delta$RA$\approx -7$\as, which creates the S1-S4 quad. But because \hres~maps have Plummer spheres with smaller widths (i.e., higher spatial resolution), they have a sharper density peak associated with that galaxy. 

The five black curves in Figure~\ref{denxparam} show publicly available reconstructions: parametric models by Sharon et al., Keeton et al., Glafic team, CATS team, and free-form hybrid model by Diego et al., along a line connecting observed SX and S1 Refsdal images. These models show less dispersion than \grale~because they assume specific functional forms for the mass distributions of the galaxy and cluster components, which artificially eliminate many degenerate mass models. \grale~reconstructions do not have such assumptions, and so \grale~curves show a fuller range of density distributions allowed by the multiple images. The realistic range of mass distributions probably lies somewhere between these two types of models: on one hand, some \grale~models are likely to be astrophysically implausible, but on the other, parametric models are too restrictive, and likely do not account for the complexity of mass distribution in lenses, especially in merging, far-from-relaxed galaxy clusters, such as MACS J1149.

In Figure~\ref{denxsec} the pink curves (left panel) show the 67 maps of the BFsubset, while the blue curves (right panel) show the 59 maps of the HRsubset. When dealing with mass maps related to each other by more than one degeneracy, and especially when these degeneracies are approximate, it can be difficult to ascribe differences in maps to specific degeneracies. Here we present what we consider to be a very reasonable interpretation, while acknowledging that other combinations of degeneracies are also possible.

As an illustration of agMSD, we highlight two \grale~maps that are closely related by it. In the left panel of Figure~\ref{denxsec}, these are $\kappa_1$ (thick solid black curve), and $\kappa_2$ (thin solid black curve). A third curve (thick dashed black) shows $\lambda\kappa_1+(1-\lambda)$, with $\lambda=0.73$, i.e., an MSD-transformed $\kappa_1$ cross-section. The cross-sections of these two maps (thin solid and thick dashed) are very similar---implying that the two are related by a degeneracy similar to MSD---but not the same, suggesting that other degeneracies, for example, MpD, could be contributing in addition to agMSD. In the right panel, the two agMSD related mass maps are $\kappa_3$ (thick solid black curve), and $\kappa_4$ (thin solid black curve), and the corresponding factor is $\lambda=0.55$. 

To demonstrate the likely presence of other degeneracies, like MpD, we first approximately take out the standard MSD, as follows. For every one of the $\kappa$ cross-sections presented in Figure~\ref{denxsec}, we find $\lambda$ such that the curve $\lambda\kappa+(1-\lambda)$ attains $\kappa=0$ at some location within the $\Delta$RA range shown in the figure. The resulting $\kappa$ cross-sections are shown in Figure~\ref{MSDdenx}. Though the maps attain $\kappa=0$ at different locations, for the vast majority it happens within a narrow interval $-0.5$\as$<\Delta$RA$<-0.3$\as. The bar code at the bottom indicates the locations of multiple images within $\pm 6$\as of the SX-S1 line, mapped on to the $\Delta$RA axis. The $\pm 6$\as range includes all but 4 of the knots of the spiral galaxy at $z=1.489$; more distant multiple images have lesser influence on the mass distribution near SX and the Refsdal quad. Because the bulk of the images in this region are knots of the same spiral galaxy, most of the agMSD is just the standard MSD and its approximate versions, arising due to small uncertainties in the image positions.

 If the maps were related by MSD transformations, all the curves would be identical. To judge the importance of agMSD one must compare the corresponding panels of Figures~\ref{denxsec} and \ref{MSDdenx}. The curves on the right side of both panels of Figure~\ref{MSDdenx}, at $\Delta$RA$\simlt -2$\as, show less dispersion compared to the corresponding panels in Figure~\ref{denxsec}, implying that agMSD is likely present in \grale~maps, and was taken out in Figure~\ref{MSDdenx}, and that the small residual differences are due to other degeneracies. For example, MpD, while still present, is confined to small regions between images, which are abundant at $\Delta$RA$\simlt -2$\as.

On the left side of the panels, at $\Delta$RA$\simgt -2$\as, where the images are sparse, MpD becomes more important, diminishing the visibility of MSD, and making the $\kappa$ cross-sections have different shapes. In other words, it is the redistribution of mass between images due to MpD that likely accounts for most of the scatter in the curves at $\Delta$RA$\simgt -2$\as~in both panels of Figures~\ref{denxsec} and \ref{MSDdenx}.

\subsection{$H_0$ from lensed images only}

We conclude that both types of degeneracies, agMSD and MpD, contribute to mass maps, and will therefore contribute to the model time delay,  $\tau_{\rm model}$, and estimation of $H_0$. 

Figure~\ref{tdxsec} shows the cross-section of the arrival time surface along the same straight lines that were used in Figures~\ref{denxsec} and ~\ref{MSDdenx}. On the vertical axis the curves were shifted to go through zero at the location of predicted SX, to make it easier to read off the model time delay between SX and S1, $\Delta \tau_{\rm model}$. The dashed curves in the two panels show 67 models of the BFsubset and 59 models of the HRsubset.

The model time delay, $\Delta \tau_{\rm model}$, and the observed time delay, $\tau_{\rm obs}$, are connected by $\Delta t_{\rm obs}=(H_{0,\rm fid}/H_0) \Delta \tau_{\rm model}$, where $H_{0,\rm fid}$ is the fiducial value of the Hubble constant assumed in \grale~runs. If the observed time delay is known, and the global cosmology is fixed, measuring $\tau_{\rm model}$ is equivalent to measuring $H_0$. In Figure~\ref{tdxsec}, $\Delta \tau_{\rm model}$ values span a factor of 10, from $\sim\!0.2$ to $\sim\!2$ years. Therefore the 123 lensed images that we use, with no additional constraints, can constrain $H_0$ only to within a factor of 10. 

The fact that images alone have any constraining power at all is due to the presence of sources at different redshifts. If all sources were at the same redshift, MSD would prevent any inference on $H_0$. But the presence of sources at a range of redshifts does not eliminate all the degeneracies, which lead to a wide range of predicted $\Delta\tau_{\rm model}$ values.

To obtain a competitive measure of $H_0$ given $\Delta t_{\rm obs}$ and cosmology, one needs to isolate a narrow range of mass models. Though lensing by itself cannot break the degeneracies, it can help us determine what additional information would be helpful. Specifically, lensing mass reconstructions can tell us what observables correlate with $\Delta \tau_{\rm model}$. 

\begin{figure}    %%   figure 9
\centering
\vspace{-15pt}
\includegraphics[width=1.0\linewidth]{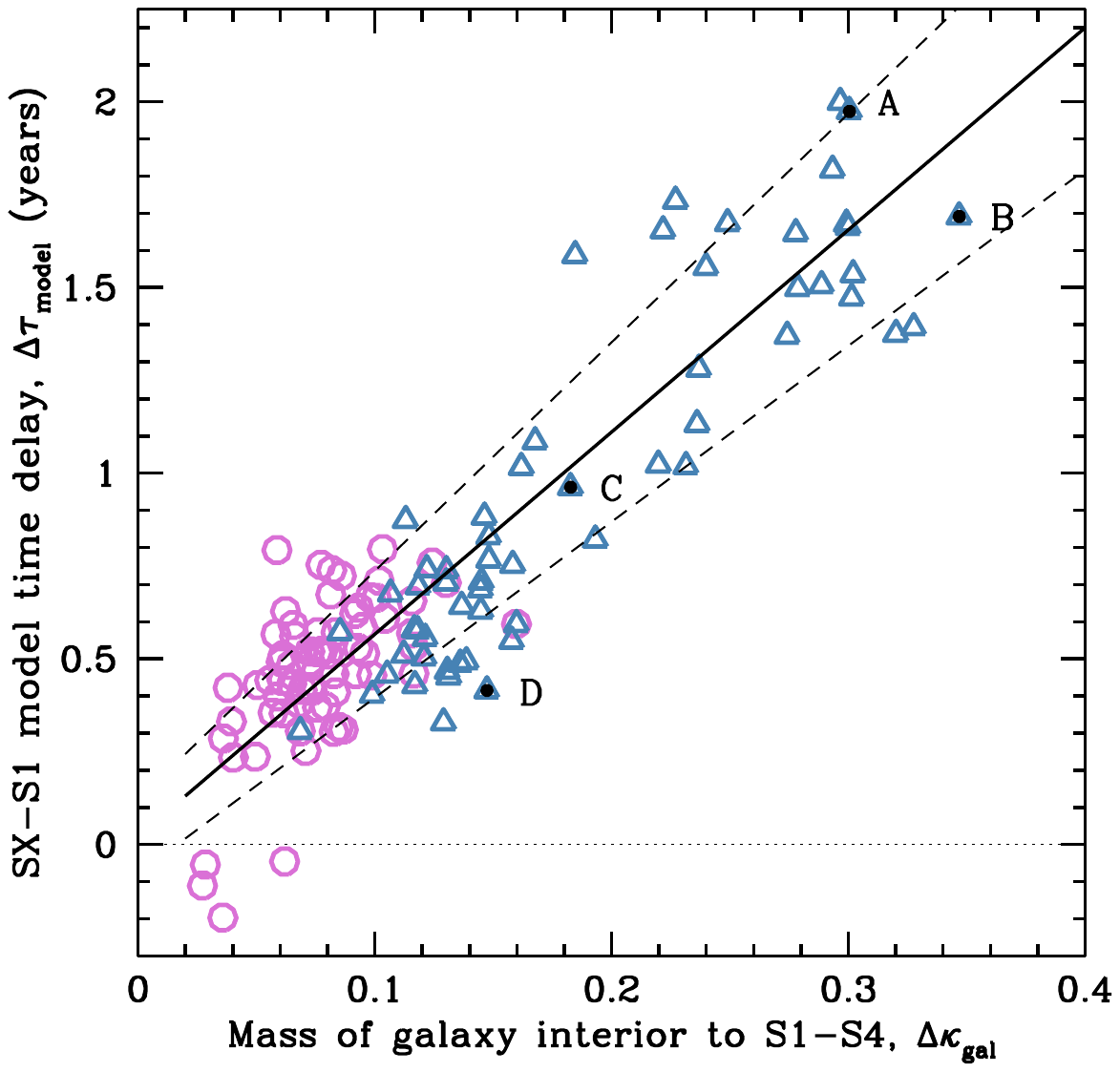}
\vspace{-65pt}
\caption{\grale~predicted time delay between SX and S1 (in years) vs. $\Delta\kappa_{\rm gal}$, which is proportional to the projected central mass of the galaxy interior to the Refsdal quad S1-S4. Pink circles (blue triangles) represent the 67 (59) BFsubset (HRsubset) maps. We measure $\Delta\kappa_{\rm gal}$ as the difference between the average densities within $r\leq 1.5$\as and $r=2.0$\as$\rightarrow 2.5$\as, where the latter is assumed to be representative of the projected density of the cluster. The thick black line is the ridge of the trend, while the thin dashed lines enclose approximately $68\%$ of all \grale~models. One-sided $1\sigma$ uncertainty ranges from $\sim 18\%$ for high galaxy masses ($\Delta\kappa_{\rm gal}=0.3$) in Figure~\ref{kappadifftd}, to $\sim 44\%$ for low galaxy masses ($\Delta\kappa_{\rm gal}=0.05$). At the galaxy mass corresponding to observed SX-S1 time delay, $\sim 1$ year \citep{kelly16a}, one-sided $1\sigma$ uncertainty is $\sim 23\%$. The mass distributions of models A, B, C and D are shown in Figure~\ref{massmap}. 
}
\label{kappadifftd}
\end{figure} 

\begin{figure*}    %%   figure 10
\centering
\vspace{-5pt}
\includegraphics[width=0.49\linewidth]{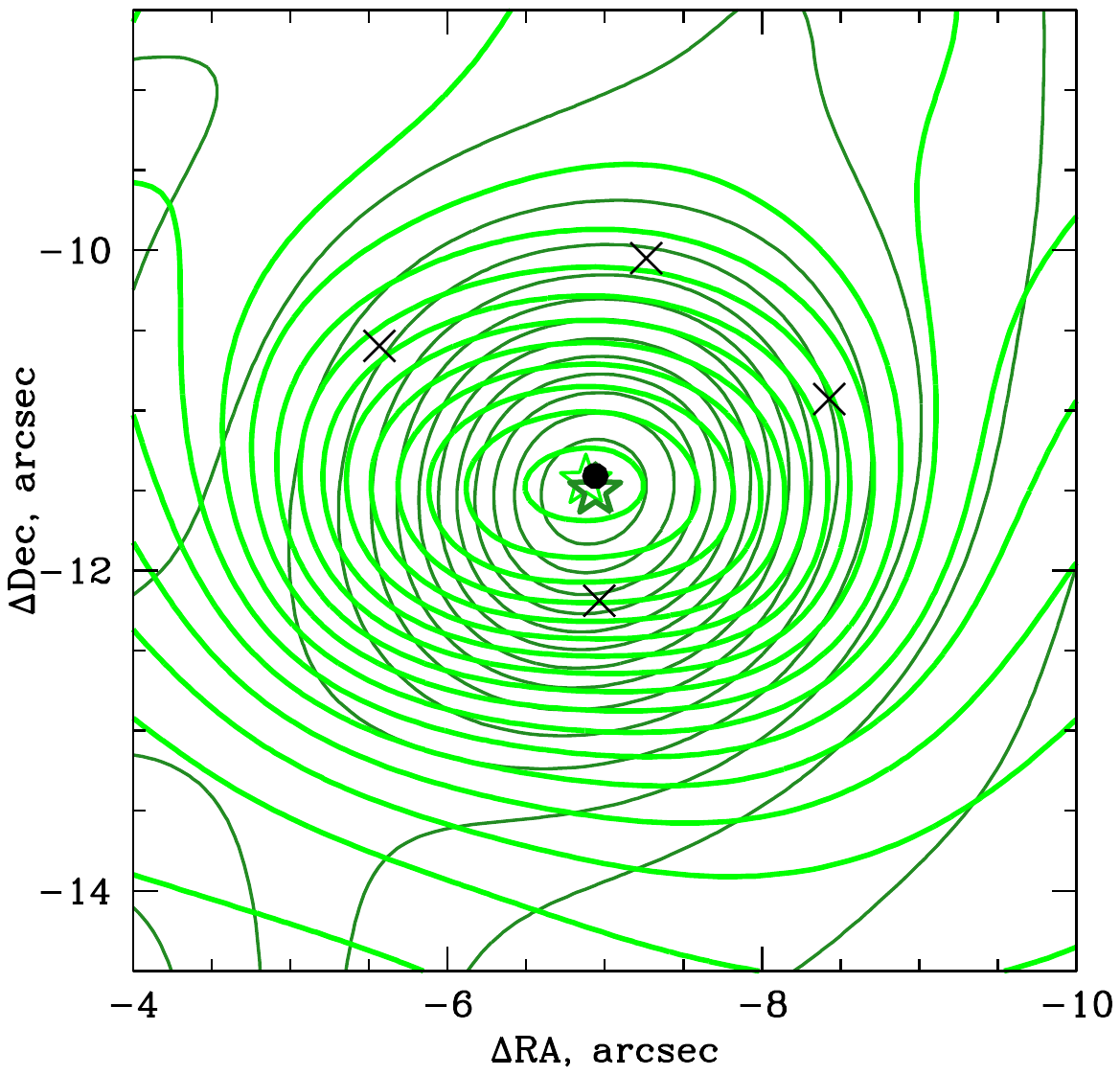}
\includegraphics[width=0.49\linewidth]{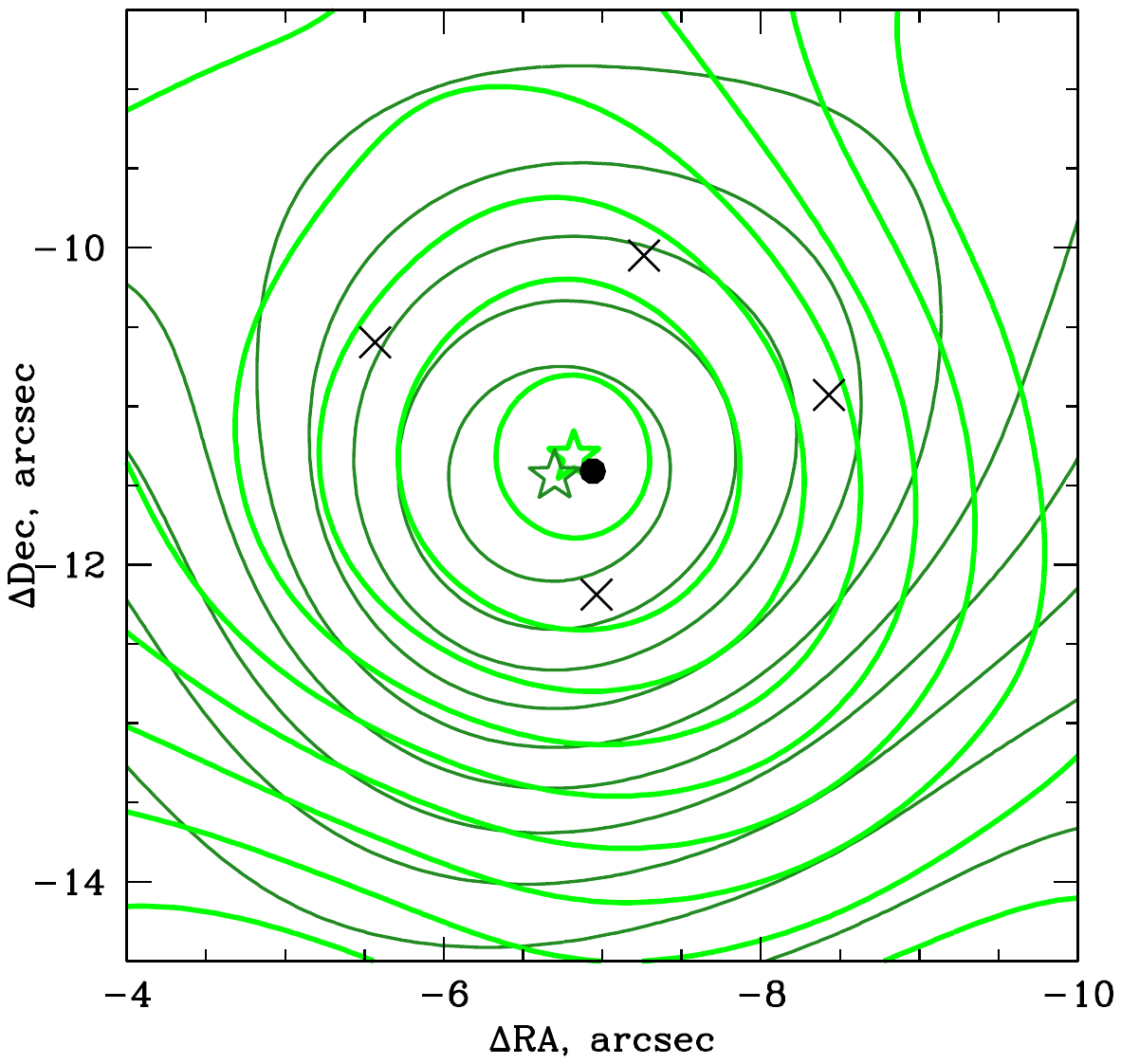}
\vspace{-50pt}
\caption{
{\it Left:} Density contours of the total mass distribution of two individual \grale~reconstructions, represented by black dots labeled A and B in Figure~\ref{kappadifftd}. The two have very similar density profile slopes near the image circle. Contours are linear and are separated by $\delta\kappa=0.05$. Four black crosses show the locations of S1-S4. The star symbols mark the centers of the two mass peaks, and the black filled circle is the center of the cluster elliptical, which splits images S1-S4. The fractional difference in the predicted $H_0$ is $\sim 22\%$.  
{\it Right:} Same as the left panel, but for the two models labeled C and D. The fractional difference in the predicted $H_0$ is $\sim 80\%$. The offset between the center of the observed galaxy and the center of the galaxy peak in the two panels ranges from 0.07\as~to 0.24\as. Note that the galaxy interior to S1-S4 was not used in the reconstruction.
}
\label{massmap}
\end{figure*} 

\subsection{$H_0$ from lensed images and galaxy masses}\label{imgal}

Figure~\ref{kappadifftd} shows that the (projected) mass of the galaxy interior to S1-S4 is correlated with model time delay, $\Delta \tau_{\rm model}$. The mass is proportional to the difference between the average $\kappa$ densities within $r\leq 1.5$\as and $r=2.0$\as$\rightarrow 2.5$\as, where the latter is assumed to be representative of the projected density of the cluster in the vicinity of the Refsdal quad. We call this quantity $\Delta\kappa_{\rm gal}$. Because the radii quoted above are the same throughout the analysis, $\Delta\kappa_{\rm gal}$ is proportional to the mass of the galaxy. The image radius is about $1.5$\as, so we are measuring the mass of the galaxy interior to the images. 

Note that $\Delta\kappa_{\rm gal}$ is equivalent to the slope of the density profile of the total (dark matter and galaxy) mass distribution. MSD, which is sometimes called the steepness transformation, changes the slope of the total mass responsible for lensing. The image separation of the S1-S4 quad fixes the total mass enclosed by the four images, but does not say how that mass is partitioned between the galaxy and the cluster's smooth dark matter at that location. Changing that partitioning affects the total density slope. In the rest of the paper we will refer to  $\Delta\kappa_{\rm gal}$ as the galaxy mass, but one must keep in mind that it is analogous to steepness.

We have tried radial ranges other than the ones specified above, but these give the strongest correlation with $\Delta \tau_{\rm model}$, i.e., the most optimistic scenario. The 67 circles and 59 triangles represent BFsubset and HRsubset, respectively. Of the 126 models, 122 have SX arriving after S1, as one would expect for the ordering of the saddle and the neighboring minimum, in a naked cusp configuration. 

The points in Figure~\ref{kappadifftd} show a well defined trend, with scatter. The main trend is primarily due to agMSD (steeper galaxy density profiles lead to larger model time delays), while the scatter is likely due to MpD, as well as the approximate nature of the agMSD, namely that the images are not reproduced exactly. 

The existence of the trend indicates that an accurate measurement of the mass of the galaxy interior to S1-S4 (which is related to the stellar velocity dispersion) will eliminate most models in Figure~\ref{kappadifftd}, and meaningfully constrain $H_0$, but the scatter in $\Delta \tau_{\rm model}$ at a given mass will remain. The thick black line shows the mean trend. The error-bars (dashed lines) were constructed as symmetric straight lines above and below the mean trend, with their slopes and intercepts adjusted so that as close to $68\%$ as possible of each of BFsubset and HRsubset models were enclosed between them. The fractional uncertainty due to scatter is not constant as a function of galaxy mass; one-sided $1\sigma$ uncertainty ranges from $\sim 18\%$ for high galaxy masses ($\Delta\kappa_{\rm gal}=0.3$) in Figure~\ref{kappadifftd}, to $\sim 44\%$ for low galaxy masses ($\Delta\kappa_{\rm gal}=0.05$). At the galaxy mass corresponding to observed SX-S1 time delay, $\sim 1$ year \citep{kelly16a}, one-sided $1\sigma$ uncertainty is $\sim 23\%$.

Figure~\ref{massmap} gives examples of two high mass (labelled A and B in Figure~\ref{kappadifftd}) and two low mass (labelled C and D) galaxies splitting S1-S4. We selected these maps such that the galaxies in each pair had similar masses; in pair A, B (C, D) the two galaxy masses differ by $\sim 20\%$ ($\sim 15\%$). The mass maps were further selected to have their galaxy centers (star symbols) be close to the center of the observed galaxy (black solid dot). In A, B (C, D) the centers are displaced by $\simlt 0.09$\as ($\simlt 0.24$\as). The isodensity contours of these mass distributions are not elliptical, but even isolated ellipticals do not have strictly elliptical isophotes \citep[e.g.,][]{hao06,mitsuda17}, and presumably isodensity contours. 

Parametric models assume fixed functional forms for the cluster and galaxy mass profiles, which artificially suppress MpD, and restrict the shape of the trend due to agMSD. Thus, under the assumptions inherent in parametric modeling, the plot equivalent to our Figure~\ref{kappadifftd} would look like a single curve, and so most, if not all degeneracies are broken by measuring the mass of the galaxy interior to S1-S4. (This is usually done using a proxy, like the central velocity dispersion.)

Since $\Delta \tau_{\rm model}$ correlates well with the mass of the galaxy interior to S1-S4, one might ask if a similarly good correlation exists between $\Delta \tau_{\rm model}$ and the mass of the galaxy, or galaxies, in the close vicinity of SX. Unfortunately, no. This is  probably related to the fact that near S1-S4, agMSD is the dominant degeneracy, and mass distributions connected by it are simple scalings of each other. On the other hand, the main degeneracy near image SX is MpD, and there is no simple relation between the mass maps connected by MpD. So the prevalence of MpD near SX means that nearby galaxy masses will not provide a useful constraint.

With the currently available set of spectroscopic lensed images, the only galaxy mass that correlates with $\Delta \tau_{\rm model}$, and hence is useful for constraining $H_0$, is the mass of the galaxy enclosed by S1-S4. If the mass of that galaxy is estimated accurately, and no parametric assumptions are used, $H_0$ can be estimated to within $\sim 23\%$. If more existing images had spec-$z$'s, or more images were discovered, it would help suppress MpD around SX and likely lead to better constraints on $H_0$.

\subsection{$H_0$ from lensed images, galaxy masses, and parametric assumptions}

The predicted time delay between S1 and SX obtained by parametric models (Jauzac et al., Sharon et al, Oguri et al., Grillo et al., Zitrin et al.) and free-form hybrid models (Diego et al.) are presented in Figure 3 of \cite{kelly16a}. Most models have $1\sigma\sim 25$ days, or $\sim 7\%$ uncertainty in $\Delta \tau_{\rm model}$. The transition from $\sim 23\%$ precision on $H_0$ to $\sim 7\%$ is accomplished by introducing parametric forms describing the distribution of mass associated with the cluster dark matter, and the individual galaxies in the cluster. Because these priors differ between various lensing inversion methods, some models---even those using the same \lenstool~software \citep{kne93,jul07}: Sharon et al. and Jauzac et al.---differ by $2\sigma-3\sigma$. This is an indication that the parametric assumptions are somewhat too restrictive; in other words, not all of these assumptions can be correct.

An eye-ball estimate of the combination of these models gives an uncertainty of $\sim 18\%$, similar to that obtained through Bayesian analysis from these data, where one finds $H_0=64^{+9}_{-11}$km~s$^{-1}$~Mpc$^{-1}$ \citep{veg18}. 

\begin{figure}    %%   figure 11
\centering
\vspace{-20pt}
\includegraphics[width=1.07\linewidth]{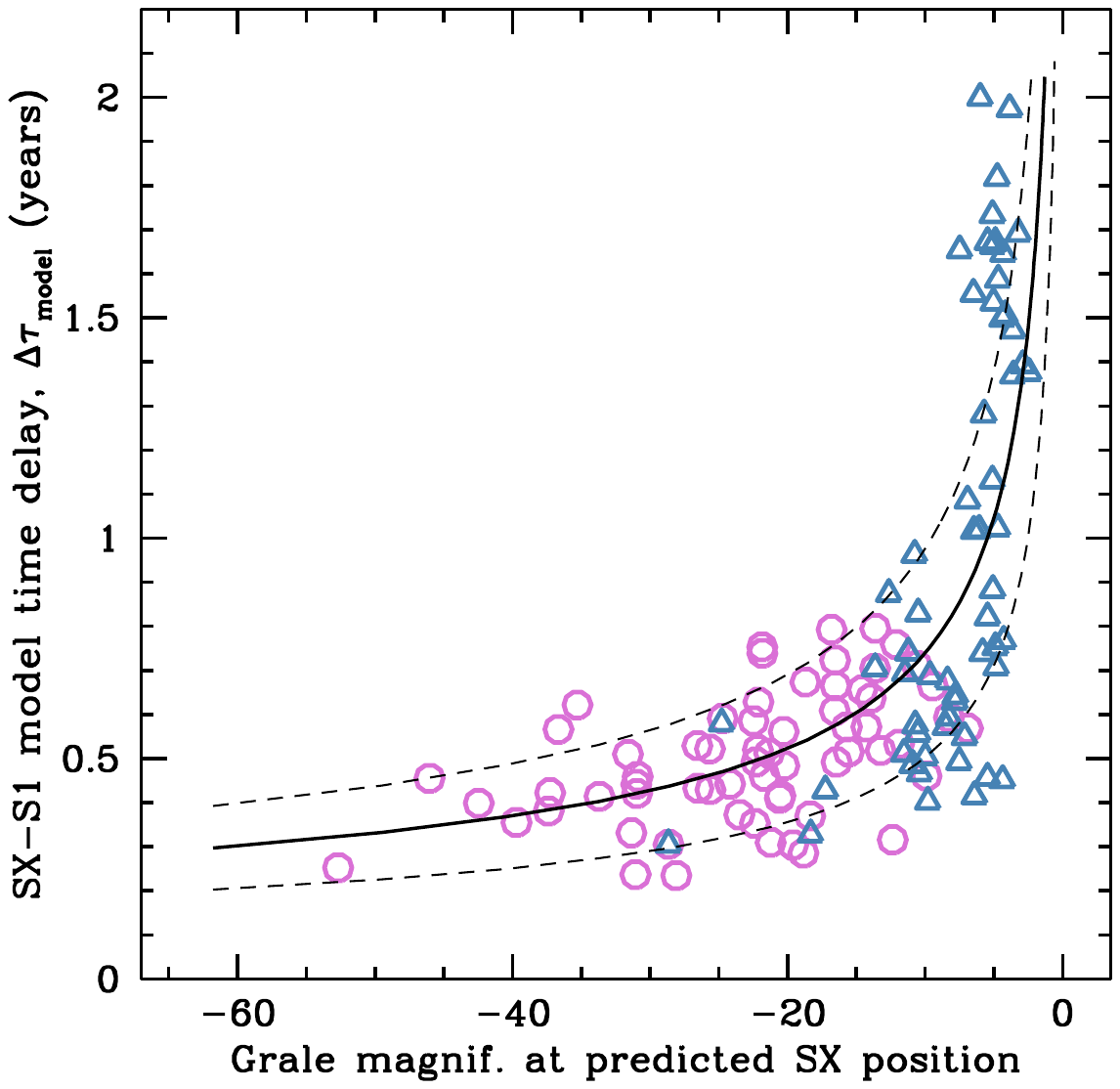}
\vspace{-65pt}
\caption{\grale~time delay vs. magnification at the predicted location of SX. (Four points from the BFubset are outside the limits of the plot.) Pink circles (blue triangles) represent the 67 (59) BFsubset (HRsubset) maps. The black thick curve was obtained using the MSD parameter, $\lambda$ (Section~\ref{illus} and Figure~\ref{MSDdenx}), as discussed in Section~\ref{magn}. The dashed lines contain approximately $68\%$ of \grale~models. If magnification at SX is know exactly, then the model time delay, and hence $H_0$ are constrained to $\pm 32\%$, at $1\sigma$.}
\label{magniftd}
\end{figure} 

\subsection{$H_0$ from lensed images and absolute magnification of SX}\label{magn}

Because the absolute fluxes of supernovae Type Ia are standardizable, they allow for precise determination of the magnification at the location of their images, and can in principle be used to break MSD in cluster lenses \citep{rod15}. Though Refsdal is not Type Ia, one can still ask if knowing the magnification will break some existing degeneracies. 

Magnification, $\mu$, is related to normalized surface mass density $\kappa$, and shear $\gamma$, through $\mu=[(1-\kappa-\gamma)(1-\kappa+\gamma)]^{-1}$. Because of MSD, model time delay is related to $\kappa$, and to $\mu$. Therefore one should expect that $\Delta\tau_{\rm model}$ between SX and S1 is related to $\mu$ at SX; this is shown in Figure~\ref{magniftd}. We will now account for the main trend, through the MSD parameter $\lambda$, used in Section~\ref{illus} and Figure~\ref{MSDdenx}, where all \grale~models have been MSD transformed to have zero mass sheet in the lens plane region between SX and the S1-S4 quad. The \grale~reconstructed density (Figure~\ref{denxsec}) and the density in Figure~\ref{MSDdenx} are related by, $\kappa_{\rm Grale}=\lambda \kappa_{\rm Fig.6}+(1-\lambda)$. Typical $\kappa$ at SX in Figure~\ref{MSDdenx} is around 1, with scatter. The black thick curve in Figure~\ref{magniftd} is obtained by assuming that $\kappa_{\rm Fig.6}=0.9$, $\gamma$ is related to $\kappa$ through $\gamma_{\rm Grale}=g(1-\kappa_{\rm Grale})$, where $g=15$, and $\Delta\tau_{\rm model}\propto l\lambda$, where $l=3.5$. (Other similar sets of values for $\kappa_{\rm Fig.6}$, $g$ and $l$ can also reproduce the main trend.) Dashed curves were obtained by scaling the mean trend symmetrically, up and down along the vertical axis, such that they enclose approximately $68\%$ of each of BFsubset and HRsubset models. This required us to displace the dashed curves from the mean trend by $\pm 32\%$. This means that knowing the exact magnification at SX translates into uncertainty in $\lambda$ of $\pm 32\%$, at $1\sigma$ level. For a known observed time delay, $H_0\propto \Delta\tau_{\rm model}$, and MSD implies that $\Delta\tau_{\rm model}\propto \lambda$, therefore $1\sigma$ uncertainty on derived $H_0$ is $\pm 32\%$, somewhat larger than what was obtained in Section~\ref{imgal} using galaxy mass. 

Can one reduce uncertainties on $H_0$ further by combining observed magnification of SX and observed mass of galaxy enclosing S1-S4?  With the current data, set this is probably unlikely; the scatter in Figures~\ref{kappadifftd} and \ref{magniftd}, which is due to MpD, prevents significant reduction in uncertainties. It appears that either galaxy mass, or magnification of SX will reduce uncertainty to approximately the same level. 

What about magnification of Refsdal image S1? Since \grale~does not accurately predict the location of the quad images, using \grale~predicted magnification---which depends on the second derivative of the lensing potential---is ill advised.

\section{Summary and Discussion}

The value of $H_0$ affects almost every aspect of cosmology, from distance scale to fundamental physics. The goal of this paper was to understand how various constraints affect the uncertainties on $H_0$ measurement based on cluster lensing.

To separate out the constraints provided by lensed images alone, we used free-form \grale, a lens inversion method that relies solely on image positions. No information about any cluster galaxy was used. \grale~predicts very accurately ($\Delta=0.017$\as$-0.036$\as) and precisely (rms~$=0.272$\as$-0.360$\as) the position of the cluster-wide saddle point, SX, and the locus of the cluster critical curve ($\simlt 0.05$\as), which goes between the transients. 

While \grale~predictions in the image plane are excellent, the predicted time delays span a very large range, implying that these lensed images alone do not constrain $H_0$ significantly.  This is due to the presence of unbroken degeneracies, mostly the approximate generalized mass sheet degeneracy (agMSD), and the monopole degeneracy (MpD). We presented an argument that their regions of dominance vary across the lens plane. In regions where multiple images at the same redshift are plentiful, agMSD likely dominates, but MpD is nearly broken. This is well exemplified by the region around Refsdal images S1-S4, where multiple knots of the background spiral abound. In the region surrounding SX, images are less plentiful, and there is no equivalent of the S1-S4 quad. Here, both agMSD and MpD play a role, and the large scatter in the shapes of density distributions in this region is probably due to MpD. Because MpD is relatively suppressed around S1-S4, and agMSD dominates, the mass of the galaxy enclosed by S1-S4 correlates well with the model time delay. Thus, measuring this galaxy's mass would effectively break agMSD, leaving just the scatter due to MpD.

We conclude that using the currently available data, the multiple images by themselves limit the range of allowable $\Delta \tau_{\rm model}$ to 0.2--2 years, or $\sim 1000\%$ uncertainty in $H_0$. Obviously, additional constraints are needed, and are, in fact, always used in $H_0$ studies. We showed that the mass of the galaxy splitting S1-S4 quad correlates well with \grale~model time delay, which is proportional to $H_0$. Including an accurate estimate of the galaxy mass would cut down on the range of mass models, and reduce uncertainties to $\sim \pm 23\%$ (at $1\sigma$), or potentially better, if image positions are reproduced very well. Absolute magnification of image SX also correlates with model time delay; its measurement would yield $\pm 32\%$ uncertainties at $1\sigma$. However, combining galaxy mass and magnification information is unlikely to improve precision significantly because the main remaining degeneracy is MpD, whose contribution to the mass distribution in the lens plane is `stochastic' and cannot be described by a scaling relation. 

With the currently available data, uncertainties smaller than $\sim 23\%$, of order a few percent, are a consequence of imposing assumptions on the shapes of the galaxy and cluster mass distributions. An interesting future exercise with \grale~would be to determine what density of images would be required to suppress MpD sufficiently, to achieve a measurement of $H_0$ with a few percent precision.

We propose that a reliable estimate of $H_0$ from Refsdal can be obtained by exploiting the advantages of two types of lensing inversion techniques: parametric, and free-form methods, like \grale, that are capable of exploring a wide range of mass models. The advantage of parametric methods is that their models incorporate our knowledge about average galaxy and cluster properties, while the advantage of \grale~and similar methods, is that they recognize that averages might not be applicable, especially in a highly non-equilibrium environment of a merging cluster like MACS J1149.

Carrying out a reliable measurement of $H_0$ would also require some additional data. It is essential to obtain an accurate estimate of the mass of the galaxy enclosed by S1-S4. Measuring spectroscopic redshifts for all known images will help improve the accuracy of the mass model \citep{joh16,wil17}. Any additional images, beyond the current set, will be also welcome. 

Given these data, free-form methods would be able to narrow down $H_0$ to $\sim 23\%$ or somewhat better, and, more importantly, provide a conservative estimate of the uncertainties \citep{pri17}. Individual parametric models will yield uncertainties of $<10\%$, or smaller, but these need not agree with each other. One way to combine the parametric models is by using the Bayesian formalism outlined in \cite{veg18}, but to weight the models (using $q_i$ parameter in eq.~7) by the lens plane rms before including them in the Bayesian analysis. A low lens plane rms is a necessary, but not a sufficient measure of how well a mass model approximates the true mass distribution. However, it remains the best available measure.

An $H_0$ measurement can be deemed reliable if \grale, or similar method, and Bayesian-combined parametric models, yield the same value to less than $\sim 2\sigma$. The combined parametric models would provide the optimistic estimate of uncertainties, while \grale, or similar, would provide the conservative uncertainties. 

\section*{Acknowledgments}

The authors are grateful to all the Hubble Frontier Fields modeling teams for providing lensed image data. The authors would like to thank Patrick Kelly and Sherry Suyu for providing useful suggestions that helped improve the paper, and Kevin Sebesta for his help with some data files. The authors acknowledge the Minnesota Supercomputing Institute for their computing time, resources, and support. The background HST image of MACS J1149 in Figure~\ref{images} was obtained from {\tt http://hubblesite.org/image/3079/category/12-cosmology.}

\bibliographystyle{mnras}
%\bibliography{bibfile} 

% Don't change these lines
\bsp	% typesetting comment
\label{lastpage}
\end{document}